\documentclass[12pt]{article}
\usepackage{putex}
\usepackage{graphicx}
\usepackage{multirow}

\def\arctanh{\mop{arctanh}}

\begin{document}

\title{Conformal hydrodynamics in Minkowski and de Sitter spacetimes}

\authors{Steven S. Gubser\worksat{\PU,}\footnote{e-mail: {\tt ssgubser@princeton.edu}} and
Amos Yarom\worksat{\PU,\Technion,}\footnote{e-mail: {\tt ayarom@princeton.edu}}}

\institution{PU}{Joseph Henry Laboratories, Princeton University, Princeton, NJ 08544}
\institution{Technion}{Department of Physics, Technion, Haifa 32000, Israel}

\abstract{We show how to generate non-trivial solutions to the conformally invariant, relativistic fluid dynamic equations by appealing to the Weyl covariance of the stress tensor.  We use this technique to show that a recently studied solution of the relativistic conformally invariant Navier-Stokes equations in four-dimensional Minkowski space can be recast as a static flow in three-dimensional de Sitter space times a line.  The simplicity of the de Sitter form of the flow enables us to consider several generalizations of it, including flows in other spacetime dimensions, second order viscous corrections, and linearized perturbations.  We also construct the anti-de Sitter dual of the original four-dimensional flow.  Finally, we discuss possible applications to nuclear physics.}

\preprint{PUPT-2358}

\maketitle
\tableofcontents

\section{Introduction}
\label{INTRODUCTION}

Analytic solutions to the Navier-Stokes equations are few and far between. In \cite{Gubser:2010ze} a new solution to the relativistic, conformally invariant Navier-Stokes equation was constructed. The solution is a generalization of Bjorken flow \cite{Bjorken:1982qr} and describes a boost-invariant medium expanding both longitudinally and radially, which makes it an attractive toy model for heavy ion collisions. We will refer to this solution as the $SO(3)\times SO(1,1) \times \mathbf{Z}_2$-invariant solution, or the $SO(3)$-invariant flow for short. The former title describes the symmetries used to construct the solution: like Bjorken flow, it is boost invariant, and it has a ${\bf Z}_2$ symmetry generated by reflections through the collision plane; but instead of being translationally invariant in the transverse directions it has an $SO(3)$ conformal symmetry. The explicit form of the inviscid $SO(3)$-invariant flow, and a simple derivation of it, can be found in section \ref{SIMPLE}.

Conformal invariance was a crucial assumption in the construction of the $SO(3)$-invariant flow.  Conformal invariance implies that the dynamics is invariant under a Weyl rescaling of the metric:
\begin{equation}
\label{Weyl}
	g_{\mu\nu} \to \Omega^{-2} g_{\mu\nu} \,,
\end{equation}
where $\Omega$ is allowed to vary across spacetime.  The Minkowski metric, the de-Sitter (dS) metric, and the anti-de Sitter (AdS) metric are all related to one another through Weyl rescalings. A Weyl rescaling locally preserves angles but not sizes. When considering the hydrodynamic evolution of the presumed quark-gluon-plasma (QGP) at RHIC and the LHC, it is a reasonable approximation in the first few ${\rm fm}/c$ to set the pressure $p$ equal to $\epsilon/3$, and to set bulk viscosity equal to $0$.  These are precisely the conditions guaranteed by the tracelessness of the stress tensor, which is the hallmark of conformal invariance.  In other words, $p=\epsilon/3$ and $\zeta=0$ means that the symmetries of relativistic hydrodynamics are enlarged to include conformal symmetry.  We discuss conformal hydrodynamics at greater length in section~\ref{CHYDRO}.

Conformal symmetry turns out to be a powerful tool for generating non-trivial solutions to the Navier-Stokes equations, both exact and approximate.  Our derivation in section~\ref{SIMPLE} of the inviscid $SO(3)$-invariant flow starts with a stationary fluid in the geometry $dS_3 \times \mathbf{R}$. Owing to the simplicity of the solution in $dS_3 \times {\bf R}$ coordinates, in section \ref{SOLUTION}, we are able to generalize it to arbitrary space-time dimensions and to include the effects of a non-vanishing chemical potential and second order viscous corrections. In section \ref{EMBEDDING} we explain how to rewrite this solution as a dual time-dependent AdS black hole geometry.

In section~\ref{ANISOTROPIES} we derive the equations that govern arbitrary linear perturbations of the $SO(3)$-invariant flow.  A key question, which we study in section~\ref{SOLUTIONS}, is whether there are instabilities.  The answer is that instabilities do exist, but for choices of parameters chosen to match approximately with RHIC collisions, the instabilities occur at times earlier than the thermalization time.  Because of the recent interest in higher order moments of the flow of real quark-gluon plasmas \cite{Alver:2010gr,Alver:2010dn,Petersen:2010cw,Qin:2010pf,Lacey:2010hw,Teaney:2010vd}, we also consider, in sections~\ref{SOLUTIONS} and~\ref{APPLICATIONS}, explicit solutions of the equations governing linear perturbations of the $SO(3)$-invariant flow.  These solutions are classified by their transformation properties under $SO(3)$.  We also explain in section~\ref{APPLICATIONS} how perturbations to the initial state can also be classified in terms of representations of $SO(3)$.

\section{A simple derivation of the $SO(3)$-invariant flow}
\label{SIMPLE}

Inviscid, conformal, relativistic hydrodynamics in $3+1$ dimensions is based on the following form for the stress tensor:
 \eqn{Tmn}{
  T_{\mu\nu} = \epsilon u_\mu u_\nu + {\epsilon \over 3} P_{\mu\nu}
 }
where $\epsilon$ is the energy density and
 \eqn{Pmn}{
  P_{\mu\nu} = u_\mu u_\nu + g_{\mu\nu}
 }
is the tensor that projects onto the local rest frame of a fluid element.  The $4$-velocity $u_\mu$ is subject to the constraint
 \eqn{um}{
  g^{\mu\nu} u_\mu u_\nu = -1 \,.
 }
The relativistic Euler equations are
 \eqn{dTmn}{
  \nabla^\mu T_{\mu\nu} = 0 \,.
 }

Consider coordinates $(\tau,\eta,x_\perp,\phi)$ for ${\bf R}^{3,1}$, such that
 \eqn{FlatSpace}{
  ds^2 = -d\tau^2 + \tau^2 d\eta^2 + dx_\perp^2 + x_\perp^2 d\phi^2 \,.
 }
This coordinate system only covers a part of ${\bf R}^{3,1}$, which is the region in the causal future of the collision plane at $\tau=0$.  We will call this region the future wedge.  A solution to equations \eno{dTmn} was studied in \cite{Gubser:2010ze} which is defined on the future wedge of ${\bf R}^{3,1}$ and takes the following form:
 \eqn{SymSoln}{
  u_\tau = -\cosh\kappa(\tau,x_\perp) \qquad
  u_\perp = \sinh\kappa(\tau,x_\perp) \qquad u_\eta = u_\phi = 0
 }
where the transverse or radial velocity is given by
 \eqn{tanhKappa}{
  v_\perp \equiv \tanh \kappa(\tau,x_\perp) = -{u_\perp \over u_\tau} = 
    {2q^2 \tau x_\perp \over 1 + q^2 \tau^2 + q^2 x_\perp^2} \,,
 }
with $q$ an arbitrary dimensionful constant with units of inverse length,
and \eqn{EpsilonFlat}{
  \epsilon = 
    {\hat\epsilon_0 \over \tau^{4/3}}
    {(2q)^{8/3} \over \left[ 1 + 2q^2 (\tau^2 + x_\perp^2) + 
      q^4 (\tau^2 - x_\perp^2)^2 \right]^{4/3}} 
 }
with $\hat{\epsilon}_0$ a dimensionless integration constant. The solution \eno{SymSoln}-\eno{EpsilonFlat} was obtained almost entirely through symmetry considerations.  However, the methods used were less than transparent because the symmetries in question are not isometries of ${\bf R}^{3,1}$, but instead conformal isometries.  The purpose of this section is to re-derive this solution using methods that make the symmetries more manifest. A more detailed exposition of our new method, including viscous corrections, can be found in sections \ref{CHYDRO} and \ref{SOLUTION}.

The key step is to make a coordinate transformation combined with a Weyl rescaling of the metric which promotes the $SO(3)$ conformal isometry to a manifest isometry.  Explicitly, the Weyl rescaling is given by
\begin{equation}
\label{TodSW}
	d\hat{s}^2 = {1 \over \tau^2}ds^2 
	={-d\tau^2 + dx_\perp^2 + x_\perp^2 d\phi^2 \over \tau^2} + d\eta^2
\end{equation}
where $ds^2$ is the line element \eno{FlatSpace}. The coordinate transformation we use is defined through the relations
 \eqn{RhoThetaCoords}{
  \sinh\rho &= -{1 - q^2 \tau^2 + q^2 x_\perp^2 \over 2q\tau} \qquad
  \tan\theta = {2q x_\perp \over 1 + q^2 \tau^2 - q^2 x_\perp^2} \,,
 }
so that the rescaled line element takes the final form
\begin{equation}
\label{TodSC}
	d\hat{s}^2 = -d\rho^2 + \cosh^2 \rho \; (d\theta^2 + \sin^2 \theta d\phi^2) + d\eta^2\,.
\end{equation}
We emphasize that the metric $d\hat{s}^2$ is not the standard flat metric on ${\bf R}^{3,1}$ because of the overall factor of $1/\tau^2$ relative to $ds^2$.  As it turns out, $d\hat{s}^2$ is the metric of $dS_3 \times {\bf R}$.  In \eqref{TodSC} we have passed to standard global coordinates on $dS_3$.  While $\phi$ in \eno{RhoThetaCoords} is the same angle as in \eno{FlatSpace} (i.e.~the angle around the beam-line), $\theta$ in \eno{RhoThetaCoords} has nothing to do with the polar angle $\vartheta = \sin^{-1} \tanh\eta$ which vanishes at mid-rapidity; instead, as \eno{RhoThetaCoords} shows, $(\rho,\theta)$ provide an alternative parametrization of the directions $(\tau,x_\perp)$.

The $(\rho,\theta,\phi,\eta)$ coordinate system have symmetries that were not manifest before: namely, the rotations of the sphere parametrized by $(\theta,\phi)$.  These symmetries comprise the promised $SO(3)$.  Its generators are isometries of $dS_3 \times {\bf R}$, but (aside from rotations in the $\phi$ direction) only conformal isometries of ${\bf R}^{3,1}$.  It is obvious how to construct an $SO(3)$-invariant velocity profile in the $(\rho,\theta,\phi,\eta)$ coordinates:
 \eqn{uhat}{
  \hat{u}_\rho = -1 \qquad \hat{u}_\theta = \hat{u}_\phi = \hat{u}_\eta = 0 \,.
 }
Clearly, there is no reference to any specific dynamics in \eno{uhat}: we are just describing a fluid which is at rest in the $dS_3 \times {\bf R}$ geometry.  To get back to a velocity profile in the future wedge of ${\bf R}^{3,1}$, we need to go back from the $(\rho,\theta,\phi,\eta)$ coordinate system to the $(\tau,x_{\perp},\phi,\eta)$ coordinate system and carry out an appropriate Weyl rescaling of the velocity field:
 \eqn{uTransforms}{
  u_\mu = \tau {\partial \hat{x}^\nu \over \partial x^\mu} \hat{u}_\nu \,,
 }
where $\hat{x}^\mu = (\rho,\theta,\phi,\eta)$ and $x^\mu = (\tau,\eta,x_\perp,\phi)$.  The explicit factor of $\tau$ on the right hand side of \eno{uTransforms} follows from the unit norm constraint,
\begin{equation}
\label{E:uconstraint}
	\hat{u}^{\mu}\hat{u}^{\nu}\hat{g}_{\mu\nu} = -1 = {u}^{\mu} {u}^{\nu} {g}_{\mu\nu}\,.
\end{equation}
Now we can recover \eqref{tanhKappa} by plugging \eno{uhat} into \eno{uTransforms}:
 \eqn{uRatio}{
  v_\perp = -{u_\perp \over u_\tau} = 
   -{\left( {\partial \rho \over \partial x_\perp} \right)_\tau \over
      \left( {\partial \rho \over \partial \tau} \right)_{x_\perp}} \,.
 }
The partial derivatives in \eno{uRatio} are to be computed starting from the defining equations \eno{RhoThetaCoords}.  Note that, in common with Bjorken flow, it is assumed that $u_\phi = u_\eta = 0$, i.e.~there is rotational invariance around the beam-line and also boost invariance.

We can go further and obtain the energy density of the $SO(3)$-invariant flow as follows.  In the $dS_3 \times {\bf R}$ frame, the energy density $\hat\epsilon$ must be a function only of de Sitter time $\rho$.  In the absence of a chemical potential the equation of state implies that the entropy density, $\hat{s}$, is proportional to $\hat\epsilon^{3/4}$.  At some fixed de Sitter time $\rho$, the total entropy per unit rapidity is given by $(4 \pi \cosh^2 \rho) \hat{s}$: this product is the entropy density times the volume of the $S^2$ parameterized by $(\theta, \phi)$. Because there is no viscosity, the total entropy can't change with $\rho$.  Thus, we conclude that
 \eqn{dsZero}{
  {d \over d\rho} \left( \hat\epsilon^{3/4} \cosh^2 \rho \right) = 0 \,,
 }
whose solution is
 \eqn{FoundEpsilon}{
  \hat\epsilon = \hat\epsilon_0 (\cosh\rho)^{-8/3} \,,
 }
where $\hat\epsilon_0$ is a constant. We use the transformation law
 \eqn{EpsilonTransform}{
  \epsilon = {\hat\epsilon \over \tau^4} \,
 }
to recover the energy density $\epsilon$ on the future wedge of flat ${\bf R}^{3,1}$.
To understand the rescaling \eno{EpsilonTransform}, recall that lengths and times in ${\bf R}^{3,1}$ are scaled by $\tau$ relative to $dS_3 \times {\bf R}$.  Volumes in ${\bf R}^{3,1}$ should therefore be scaled by $\tau^3$ relative to corresponding volumes in $dS_3 \times {\bf R}$, and energies should be scaled by $1/\tau$; thus energy density should be scaled by $1/\tau^4$, as we have done in \eno{EpsilonTransform}.  We give a more formal derivation of \eqref{EpsilonTransform} in section \ref{CHYDRO}.

In summary, the seemingly complicated solution \eno{SymSoln}-\eno{EpsilonFlat} can be recovered through a conformal transformation from isentropic evolution of a static fluid in $dS_3 \times {\bf R}$.  The simplicity of the $dS_3 \times {\bf R}$ picture invites generalizations, and we will devote the rest of this paper to exploring some of them.  Shear viscosity (already treated in \cite{Gubser:2010ze}), higher-derivative corrections to hydrodynamics, finite chemical potentials, and anisotropies are all susceptible to partially analytic treatment in this framework, as are generalizations to other dimensions.

A peculiar feature of the $dS_3 \times {\bf R}$ conformal frame should be mentioned before going further: the coordinates $(\tau,x_\perp,\phi)$ cover only half of the full $dS_3$ geometry, whereas the coordinates $(\rho,\theta,\phi)$ cover the whole.  Moreover, the half covered by the coordinates $(\tau,x_\perp,\phi)$ is a {\it contracting} geometry.  What this means is that the distance between two points at fixed coordinates in the transverse plane decreases as $\tau$ increases.  Explicitly, this distance is ${x \over \tau}$ if $x$ is the coordinate distance in the transverse plane.  In appendix~\ref{A:deSitter} we provide more detail about different ways of putting coordinates on de Sitter space, and in figure~\ref{F:hyperbola} we show how surfaces of constant $\tau$ cut across de Sitter space.

\section{Conformal transformations and hydrodynamics}
\label{CHYDRO}

Having presented in section~\ref{SIMPLE} a simple example illustrating our new method of finding solutions to relativistic conformal hydrodynamics, we aim in this section to give a more systematic summary of the conformally invariant Navier-Stokes equations, in preparation for exploration of more complicated fluid flows in later sections.  The material in this section is not new.  Much it is part of the classical literature on hydrodynamics: see for example \cite{LandL}.  More recent literature includes \cite{Baier:2007ix,Loganayagam:2008is,Bhattacharyya:2008jc,Erdmenger:2008rm}.

To fix notation, let's re-express the Weyl rescaling \eno{Weyl} as a relation between two metrics on the same $d$-dimensional spacetime:
 \eqn{ConformalEquivalence}{
  ds^2 = \Omega^2 d\hat{s}^2 \,.
 }
Recall that the conformal factor $\Omega$ is allowed to depend on the spacetime coordinates, but should be everywhere non-zero and non-singular.  Note that a Weyl rescaling is not a coordinate transformation: Indeed, we can use the same coordinates for the metrics $ds^2$ and $d\hat{s}^2$.

It will be important for us to track how various quantities transform under Weyl rescalings.  In general, we say that a quantity $X$ transforms homogeneously with weight $\alpha_X$ under a Weyl transformation if
 \eqn{XTransform}{
  X = \Omega^{-\alpha_X} \hat{X} \,.
 }
$X$ may be a scalar quantity or a tensor, and we will refer to $\alpha_X$ as its conformal weight.  We will abbreviate \eno{XTransform} as
 \eqn{Xdim}{
  [X] = \alpha_X \,.
 }
Note that the conformal weight $\alpha_X$ depends on the index structure of $X$: For example, we read off from \eno{ConformalEquivalence} that $[g_{\mu\nu}] = -2$, but the inverse metric has weight $[g^{\mu\nu}] = +2$.

Now let's discuss the conformal properties of the energy momentum tensor and a conserved charge current.  The energy momentum tensor is given by
\begin{equation}
	\langle T^{\mu\nu} \rangle = \frac{2}{\sqrt{-g}} \frac{\delta \ln Z}{\delta g_{\mu\nu}}
\end{equation}
where $Z$ is the partition function of the theory.  In a $d$ dimensional conformal theory, and in the absence of conformal anomalies,\footnote{In practice, quantum conformal theories are anomalous \cite{BandD}. However, when considering hydrodynamics as a derivative expansion this anomaly appears only at order $d$ (for even $d$). Thus, to the order we are considering in this work the conformal anomaly can be neglected. See \cite{Baier:2007ix} for details.} $Z$ is invariant under conformal transformations. Thus,
\begin{equation}
\label{E:Tmnweight}
	[T^{\mu\nu}]=d+2\,.
\end{equation}
One can check that \eqref{E:Tmnweight} together with tracelessness of the stress tensor implies that energy-momentum conservation is a conformal-frame independent statement. A similar relation can be established for a conserved current $J^{\mu}$: If we want the conservation equation
\begin{equation}
\label{E:JConservation}
	\nabla_{\mu} J^{\mu} = 0
\end{equation}
to be conformal-frame independent, then we need
 \eqn{Jdimension}{
  [J^{\mu}]=d \,.
 }

In the hydrodynamic approximation, the degrees of freedom of the theory reduce to the energy density $\epsilon$, a velocity field $u^{\mu}$ normalized such that $u^{\mu}u_{\mu}=-1$, and charge densities $n_i$ (which one can substitute for chemical potentials, depending on the choice of ensemble).\footnote{In the case of a spontaneously broken symmetry the gradient of the Goldstone mode is also a hydrodynamic degree of freedom and can be interpreted as a superfluid velocity.} For the sake of clarity we will consider turning on only one type of charge, $n$. The index $\mu$ runs from $0$ to $d-1$. Since $g_{\mu\nu}u^{\mu}u^{\nu}=-1$ we find that $[u^{\mu}]=1$. Further, in a frame where $u_{\mu}T^{\mu\nu}=-\epsilon u_{\nu}$ we can write
\begin{equation}
	T^{\mu\nu} = \epsilon u^{\mu}u^{\nu} + T_{\bot}^{\mu\nu}
\end{equation}
with $u_{\mu}T_{\bot}^{\mu\nu}=0$. Hence, $[\epsilon] = d$. Similarly one can show that $[n] = d-1$. Consequently $[T] = [\mu] = 1$ where $T$ is the temperature and $\mu$ is the chemical potential.  The conformal weights of quantities to be encountered in the rest of this paper are summarized table~\ref{ConformalWeights}.
\newcommand\nextline{\rule[-1.2ex]{0pt}{3.7ex} \\}
 \begin{table}
  \begin{center}
  \begin{tabular}{| c | c | c | l |}
   \hline
   quantity & conformal weight & description & extra relations \nextline \hline\hline
   $g_{\mu\nu}$ & $-2$ & metric & $ds^2 = g_{\mu\nu} dx^\mu dx^\nu$  \nextline \hline
   $u_\mu$ & $-1$ & velocity & $g^{\mu\nu} u_\mu u_\nu = -1$  \nextline \hline
   \multirow{2}{*}{$P_{\mu\nu}$} & \multirow{2}{*}{$-2$} & projects onto & 
     \multirow{2}{*}{$P_{\mu\nu} = u_\mu u_\nu + g_{\mu\nu}$}  \nextline
     & & local rest frame &  \\[5pt] \hline
   \multirow{2}{*}{$T_{\mu\nu}$} & \multirow{2}{*}{$d-2$} & 
    \multirow{2}{*}{stress-energy tensor} &
     $T^\mu_\mu = 0$, $\nabla^\mu T_{\mu\nu} = 0$, \nextline
    & & & $\langle T^{\mu\nu} \rangle = \displaystyle{2 \over \sqrt{-g}} {\delta \ln Z \over \delta g_{\mu\nu}}$
      \\[10pt] \hline
   $\epsilon$ & $d$ & energy density & $T_{\mu\nu} = \epsilon u_\mu u_\nu + p P_{\mu\nu} + \ldots$  
   \nextline \hline
   $p$ & $d$ & pressure & $p = \displaystyle{\epsilon \over d-1}$ \rule[-2.2ex]{0pt}{5.5ex} \\ \hline
   $T$ & $1$ & temperature & $\epsilon \propto T^d$ if $\mu=0$  \nextline \hline
   $s_\mu$ & $d-2$ & entropy current & $\nabla^\mu s_\mu \geq 0$, $s_\mu = s u_\mu + \ldots$  \nextline \hline
   $J_\mu$ & $d-2$ & conserved current & $\nabla^\mu J_\mu = 0$,
     $\langle J^\mu \rangle = \displaystyle{{1 \over \sqrt{-g}} {\delta\ln Z \over \delta A^\mu}}$ 
      \rule[-3ex]{0pt}{7ex} \\ \hline
   $n$ & $d-1$ & charge density & $J_\mu = n u_\mu + \ldots$ \nextline \hline
   $\mu$ & $1$ & chemical potential & $\epsilon + p = Ts + \mu n$ \nextline \hline
  \end{tabular}
  \end{center}
  \caption{Conformal weights and additional relations for tensors of interest. The additional relations are valid provided the theory is free of anomalies and there are no Goldstone modes. Omitted terms, denoted $\ldots$, are higher derivative corrections.  Lorentz indices are raised using $g^{\mu\nu}$, which changes the conformal weight by $+2$.}\label{ConformalWeights}
 \end{table}

The transverse component of the energy momentum tensor can be expanded in a gradient expansion.  For a conformal theory one finds that to first order in gradients,
\begin{equation}
\label{E:Tmn0and1}
	T^{\mu\nu} = \epsilon u^{\mu}u^{\nu} + \frac{\epsilon}{d-1} P^{\mu\nu} -  \eta \sigma^{\mu\nu} \end{equation}
where the projection matrix $P^{\mu\nu}$ was defined in \eqref{Pmn} and
\begin{equation}
	\sigma_{\mu\nu} = 2 \nabla_{\langle\mu}u_{\nu\rangle} \,.
\end{equation}
Here, brackets denote a symmetric traceless projection onto the space orthogonal to $u^{\mu}$.  More explicitly,
\begin{equation}\label{Brackets}
	A_{\langle \mu\nu \rangle} = \frac{1}{2} P_{\mu \alpha}P_{\nu\beta} \left(A^{\alpha \beta} + A^{\beta \alpha}\right) - \frac{1}{d-1} P_{\mu\nu}P^{\alpha\beta} A_{\alpha \beta}\,.
\end{equation}

Similarly, in the presence of a single conserved charge $n$, and in the absence of anomalies, the associated conserved current $J$ is given by
\begin{equation}
\label{E:CurrentExpansion}
	J_{\mu} = n u_{\mu} - \kappa P_{\mu\nu} \partial^{\nu} \frac{\mu}{T}\,.
\end{equation}
The relativistic version of the Navier-Stokes equation are simply energy-momentum conservation:
\begin{equation}
\label{E:TmnConservation}
	\nabla_{\mu}T^{\mu\nu} = 0\,.
\end{equation}
In the presence of charges one must also impose \eqref{E:JConservation}.

\section{Solutions with $SO(d-1) \times SO(1,1) \times {\bf Z}_2$ symmetry}
\label{SOLUTION}

Consider a conformal theory on ${\bf R}^{d-1,1}$.  The Minkowski metric $\eta$ has $d(d-1)/2$ Killing vectors and $d+1$ additional conformal Killing vectors.  As a solution generating technique, an obvious strategy is to impose symmetry under a subset of these isometries and/or conformal isometries.  In order to reduce the Navier-Stokes equations from partial differential equations (PDEs) with $d$ independent variables (namely the coordinates on ${\bf R}^{d-1,1}$) to ordinary differential equations (ODEs), one needs to impose $d-1$ independent symmetry constraints.  This is exactly what is done in Bjorken flow: starting with ${\bf R}^{3,1}$, one needs $3$ symmetry constraints to get ODEs, and the independent symmetries used are boost invariance, rotational invariance around the beamline, and translation in one of the two directions transverse to the beamline.  Invariance under translations in the second transverse dimension follows, because the generator of such translations is the commutator of rotations and translations in the first transverse direction.  Similarly, the flow explained in section~\ref{SIMPLE} has three independent continuous symmetries: one for $SO(1,1)$ boost invariance plus two for $SO(3)$ conformal symmetry, with symmetry under the third generator of $SO(3)$ following from commutations of the other two generators.

The plan of this section is to seek elementary generalizations of the $SO(3)$-invariant flow explained in section~\ref{SIMPLE} to arbitrary dimension and non-zero shear viscosity.  We will use the same trick of passing via a Weyl rescaling from the future wedge of ${\bf R}^{d-1,1}$ to $dS_{d-1} \times {\bf R}$, and then (for the most part) consider fluid flows that are stationary in $dS_{d-1} \times {\bf R}$.  By assumption, our flows possess $SO(d-1)$ symmetry (which acts on the $S^{d-2}$ spatial slices of $dS_{d-1}$), $SO(1,1)$ boost symmetry (which acts additively on the ${\bf R}$ factor of the geometry, parametrized by rapidity $\eta$), and a ${\bf Z}_2$ symmetry under $\eta \to -\eta$.  The way the counting of the previous paragraph works out is that $SO(d-1)$ symmetry can be expressed as $d-2$ independent constraints, while $SO(1,1)$ imposes one more independent constraint; thus we have precisely the $d-1$ independent constraints we need to reduce the $d$-dimensional Navier-Stokes equations to PDEs.

\subsection{Switching from ${\mathbf{R}}^{d-1,1}$ to $dS_{d-1} \times {\mathbf{R}}$}
\label{SWITCHING}

Let's start by passing from Cartesian to Bjorken coordinates on ${\bf R}^{d-1}$:
 \eqn{BjorkenCoords}{
  ds^2 &= -dt^2 + dz^2 + d\vec{x}_\perp^2  \cr
    &= -d\tau^2 + \tau^2 d\eta^2 + d\vec{x}_\perp^2 \,,
 }
where $\vec{x}_\perp = (x^1,x^2,\ldots,x^{d-2})$ parametrizes the transverse plane ${\bf R}^{d-2}$, and $\tau$ and $\eta$ are related to $t$ and $z$ through
\begin{equation}
	\tau = \sqrt{z^2 - t^2} \qquad \tanh \eta = \frac{z}{t}\,.
\end{equation}
The region $\tau>0$ covers the future wedge $t>|z|$. In this coordinate system the $SO(1,1)$ symmetry manifests itself as translations in $\eta$. 

We want to switch to the $dS_{d-1} \times {\bf R}$ conformal frame, but first we want to provide a bit more detail about $dS_{d-1}$.  Its line element is 
\begin{equation}
\label{E:dSPatch}
	d\tilde{s}^2 = \frac{1}{\tau^{2}} \left(-d\tau^2 + d \vec{x}_{\perp}^2 \right) \,.
\end{equation}
It is helpful to recall that $dS_{d-1}$ can be realized as the locus of points in ${\bf R}^{d-1,1}$ satisfying the equation
 \eqn{dSembedding}{
  -(X^0)^2 + \sum_{m=1}^{d-1} (X^m)^2 = 1 \,.
 }
If we parameterize this surface by
 \eqn{PoincareToCovering}{
  X^0 = -{1 - q^2 \tau^2 + q^2 x_\perp^2 \over 2q\tau} \qquad
  X^i = {x_\perp^i \over \tau} \qquad
  X^{d-1} = {1 + q^2 \tau^2 - q^2 x_\perp^2 \over 2q\tau} \,,
 }
where $q$ is an arbitrary parameter,  then \eqref{E:dSPatch} can be recovered as the natural metric on the locus \eno{dSembedding} inherited from the standard flat metric on ${\bf R}^{d-1,1}$.  On the other hand, if we express an arbitrary point on $S^{d-2}$ as a unit vector $r^m$ in ${\bf R}^{d-1}$, then we may relate
 \eqn{GlobalToCovering}{
  X^0 = \sinh \rho \qquad X^m = \cosh \rho \; r^m \,.
 }
with a resulting line element for $dS_{d-1}$:
\begin{equation}
\label{E:dSGlobal}
	d\tilde{s}^2 =  -d\rho^2 + \cosh^2 \rho \; d\Omega_{d-2}^2
\end{equation}
where $d\Omega_{d-2}^2$ is the metric of the $d-2$-dimensional unit sphere. Note that $\rho$ is a time coordinate: it is global time in $dS_{d-1}$.\footnote{In comparing to \cite{Gubser:2010ze}, it is useful to note that $g = -\sinh\rho$.}

Evidently, the future wedge of ${\bf R}^{d-1,1}$ is conformally equivalent to $dS_{d-1} \times {\bf R}$:
\eqn{NewFrames}{
  ds^2 &= \tau^2 \left[ {-d\tau^2 + d\vec{x}_\perp^2 \over \tau^2} + d\eta^2 \right]  \cr
   &= \tau^2 \left[ -d\rho^2 + \cosh^2 \rho \; d\Omega_{d-2}^2 + d\eta^2 \right]  \,.
 }
We note in passing that the future wedge of ${\bf R}^{d-1,1}$ is also conformally equivalent to $AdS_2\times S^{d-2}$: The Minkowski metric on ${\bf R}^{d-1,1}$ may be expressed as
 \eqn{AdSFrame}{
  ds^2 = \tau^2 \cosh^2 \rho \left[ {-d\rho^2 + d\eta^2 \over \cosh^2 \rho} + 
    d\Omega_{d-2}^2 \right] \,.
 }
In order to present explicit formulas for how hydrodynamic quantities transform in passing from ${\bf R}^{d-1,1}$ to $dS_{d-1} \times {\bf R}$, it is useful first to parametrize the transverse plane ${\bf R}^{d-2}$ in polar coordinates, so that the standard metric on the future wedge of ${\bf R}^{3,1}$ can be written as
 \eqn{PolarCoords}{
  ds^2 = -d\tau^2 + \tau^2 d\eta^2 + dx_\perp^2 + x_\perp^2 d\Omega_{d-3}^2 \,,
 }
where $d\Omega_{d-3}^2$ is the metric on the unit sphere $S^{d-3}$, parametrized (in a manner we need not specify precisely for present purposes) by $d-3$ angles $\phi_i$.  From here on, our preferred coordinates for ${\bf R}^{d-1,1}$ will be
 \eqn{xmuPreferred}{
  x^\mu = (\tau,x_\perp,\phi_1,\ldots,\phi_{d-3},\eta) \,.
 }
On the other hand, we may express the metric on $dS_{d-1} \times {\bf R}$ as
 \eqn{PolardS}{
  d\hat{s}^2 = -d\rho^2 + \cosh^2 \rho \left( d\theta^2 + \sin^2 \theta \, d\Omega_{d-3}^2
    \right) + d\eta^2 \,,
 }
and our preferred coordinates on $dS_{d-1} \times {\bf R}$ will be
 \eqn{xhatmuPreferred}{
  \hat{x}^\mu = (\rho,\theta,\phi_1,\ldots,\phi_{d-3},\eta) \,.
 }
The map we explained from the future wedge of ${\bf R}^{d-1,1}$ to $dS_{d-1} \times {\bf R}$ in equations (\ref{BjorkenCoords})-(\ref{NewFrames}) leaves the $\phi_i$ coordinates alone.  The map sends $(\tau,x_\perp) \to (\rho,\theta)$ through precisely the formulas \eno{RhoThetaCoords} introduced in section~\ref{SIMPLE}.  The energy density, charge density, and velocity fields transform as follows:
 \eqn{TransformHydro}{
  \epsilon &= \tau^{-d} \hat{\epsilon}  \cr
   n & = \tau^{-(d-1)} \hat{n} \cr
  u_\tau &=\tau \left( {\partial\rho \over \partial\tau} \hat{u}_\rho + 
    {\partial\theta \over \partial\tau} \hat{u}_\theta \right)  \cr
  u_\perp &= \tau \left( {\partial\rho \over \partial x_\perp} \hat{u}_\rho + 
    {\partial\theta \over \partial x_\perp} \hat{u}_\theta \right)  \cr
  u_{\phi_i} &= \tau \hat{u}_{\phi_i}  \cr
  u_\eta &= \tau \hat{u}_\eta \,
 }
where we consistently use hats to denote quantities in the $dS_{d-1} \times {\bf R}$ conformal frame, while unhatted quantities are for ${\bf R}^{d-1,1}$.

In the rest of this paper, we will work almost exclusively in the $dS_{d-1} \times {\bf R}$ conformal frame, as it is understood from \eno{TransformHydro} how to go back to ${\bf R}^{d-1,1}$ quantities.

\subsection{The inviscid case}
\label{INVISCID}

Having chosen the $dS_{d-1} \times {\bf R}$ conformal frame, parametrized by variables $\hat{x}^\mu = (\rho,\theta$, $\phi_1,\ldots,\phi_{d-3},\eta)$, let us now consider the velocity profiles $\hat{u}_\mu$ permitted by $SO(d-1) \times SO(1,1) \times {\bf Z}_2$ symmetry.  The $SO(1,1) \times {\bf Z}_2$ symmetry implies $\hat{u}_\eta = 0$ and prevents the other components of $\hat{u}_\mu$ from depending on $\eta$.  For $d>3$, the $SO(d-1)$ dependence, on top of $SO(1,1) \times {\bf Z}_2$, leaves only one possible velocity profile:
 \eqn{uChoice}{
  \hat{u}_\rho = -1
 }
with all other components set to $0$.\footnote{The sign in \eno{uChoice} is fixed by noting from the first line of \eno{RhoThetaCoords} that $\tau \to 0$ corresponds to $\rho \to -\infty$.  Thus the future direction is toward more positive $\tau$ and toward more positive $\rho$.} When $d=3$ the $u_\theta$ component of the velocity profile does not necessarily vanish. We will treat the $d=3$ case with non-vanishing angular momentum in the $\theta$ direction separately in section \ref{MOMENTUM}.

Before working out the energy density, let's consider how charge density evolves.  Without appealing to the explicit form \eno{E:CurrentExpansion} of the conserved charge current $\hat{J}^\mu$, we note that the $SO(d-1) \times SO(1,1) \times {\bf Z}_2$ forces all components of $\hat{J}^\mu$ to vanish except 
 \eqn{hatJeq}{
  \hat{J}^\rho = -\hat{u}_\mu \hat{J}^\mu \equiv \hat{n} \,.
 }
Plugging \eno{hatJeq} into the conservation equation \eno{E:JConservation} leads immediately to
 \eqn{nEvolve}{
  {d \over d\rho} \left[ (\cosh\rho)^{d-2} \hat{n} \right] = 0 \,,
 }
whose general solution is 
\begin{equation}
 \label{E:invisciddensity}
	\hat{n} = \hat{n}_0 (\cosh\rho)^{-(d-2)}
\end{equation}
where $\hat{n}_0$ is a constant.  Note that the result \eno{E:invisciddensity} has no dependence on the diffusion constant $\kappa$, which makes sense because no net diffusion occurs during the flow.
 
So far we have made no appeal to dynamical equations of motion: in particular, our choice of velocity field \eno{uChoice} should be valid equally for the inviscid and viscous cases, and the result \eno{E:invisciddensity} requires only symmetry considerations plus the conservation law for charge.  Therefore, all the analysis so far is valid equally in the inviscid and viscous cases.

In order to determine the energy density for the inviscid case, we can use the same style of argument as in section~\ref{SIMPLE}.  Entropy is not produced in the absence of viscosity, so the entropy density $\hat{s}$ in $dS_{d-1} \times {\bf R}$ behaves just like the conserved charge $\hat{n}$: that is, $s \propto (\cosh\rho)^{-(d-2)}$.  Conformal invariance requires that the energy density $\hat\epsilon$ in $dS_{d-1} \times {\bf R}$ obeys an equation of state of the form
 \eqn{epsilonConformal}{
  \hat\epsilon = \hat{n}^{d \over d-1} \tilde\epsilon(\sigma)
 }
where
 \eqn{sigmaDef}{
  \sigma = {\hat{s} \over \hat{n}} \,.
 }
Conformal invariance tells us nothing about the scaling function $\tilde\epsilon(\sigma)$.  The crucial point is that $\sigma$ is constant throughout the flow, simply because $\hat{n}$ and $\hat{s}$ have the same dependence on $\rho$ (up to overall constants) in the absence of viscosity.  Thus $\hat\epsilon \propto \hat{n}^{d \over d-1}$ and combining this result with \eno{E:invisciddensity} we find
\eqn{FoundEpsilond}{
	\hat\epsilon = \hat\epsilon_0 (\cosh\rho)^{-{d(d-2) \over d-1}} \,,
}
where $\hat{\epsilon}_0$ is a constant.

\subsection{Viscous corrections}
\label{VISCOUS}

In section~\ref{INVISCID}, we saw that the solution of conformal relativistic hydrodynamics respecting $SO(d-1) \times SO(1,1) \times {\bf Z}_2$ symmetry is essentially trivial, even for the most general equation of state \eno{epsilonConformal} allowed by conformal invariance in the presence of non-zero charge density.  The reason for this triviality is that the evolution of entropy density in de Sitter time $\rho$ exactly tracks the evolution of the charge density.  Viscosity will spoil this.  In order to make progress, we need a clever parametrization of shear viscosity:
 \eqn{etaExpress}{
  \hat\eta = \hat{n} \, \tilde\epsilon'(\sigma) \tilde\eta(\sigma) \,,
 }
where $\tilde\eta$ is another dimensionless function of $\sigma$.  Because $\tilde\eta$ is arbitrary, the form \eno{etaExpress} is the most general one allowed by conformal invariance.  It seems that it would be simpler to omit the factor $\tilde\epsilon'(\sigma)$, but including it is convenient because then $\tilde\epsilon(\sigma)$ factors out completely from the hydrodynamic equations for the stationary fluid in $dS_{d-1} \times {\bf R}$.  These equations boil down to
 \eqn{SigmaForm}{
  {1 \over \tilde\eta(\sigma)} 
    {d\sigma \over d\rho} = {2(d-2) \over d-1} n^{-{1 \over d-1}} \tanh^2 \rho \,,
 }
which is separable.  A slight additional generalization is helpful: we need not insist that $\sigma$ is precisely the ratio $\hat{s}/\hat{n}$; instead, it can be some function of this ratio without changing any aspect of the calculations we have done, provided we employ the parametrizations \eno{epsilonConformal} and \eno{etaExpress}.

To proceed further, let us define
 \eqn{HyperFunction}{
  F_d(\rho) \equiv \int_0^\rho dr \, (\cosh r)^{d-2 \over d-1} \tanh^2 r =
   {\sinh^3 \rho \over 3} {}_2F_1\left( {3 \over 2},{2d-1 \over 2d-2};
    {5 \over 2}; -\sinh^2 \rho \right) \,.
 }
For any fixed $d$, the map $\rho \to F_d(\rho)$ is a smooth bijection of ${\bf R}$ to ${\bf R}$, and its slope is positive except at $\rho=0$.  Moreover, it is odd under $\rho \to -\rho$.  
The general solution of \eno{SigmaForm} can be readily expressed in terms of $F_d$:
 \eqn{GeneralSigma}{
  \int_{\sigma_0}^\sigma {du \over \tilde\eta(u)} = {2(d-2) \over d-1} \hat{n}_0^{-{1 \over d-1}}
    F_d(\rho) \,,
 }
where $\sigma_0$ is the value of $\sigma$ at de Sitter time $\rho=0$.

The solution \eno{GeneralSigma} is slightly abstract.  Let us work out an example to show its utility.  Consider the case where neither the energy density nor the viscosity depend significantly on the conserved charge.  Then we define
 \eqn{epsilonExample}{
  \hat\epsilon = \hat{T}^d \qquad\hbox{and}\qquad
   \hat\eta = {\rm H}_0 \hat{T}^{d-1} \,,
 }
where $\hat{T}$ is proportional to the temperature, and ${\rm H}_0$ is a  dimensionless constant.  ${\rm H}_0$ must remain arbitrary if we want to capture the dynamics of arbitrarily viscous fluids.  The conserved current is now treated in a probe approximation: \eno{E:invisciddensity} still holds, but by assumption, $\hat{n}$ does not enter into the stress tensor in any way.  We may nevertheless employ the parametrizations \eno{epsilonConformal} and \eno{etaExpress}, for example with
 \eqn{sigmaChoice}{
  \tilde\epsilon(\sigma) = \sigma^d \qquad\qquad
  \tilde\eta(\sigma) = {{\rm H}_0 \over d} \qquad\qquad
  \sigma = {\hat{T} \over \hat{n}^{1 \over d-1}} \,.
 }
Then it is easy to see that \eno{GeneralSigma} becomes
 \eqn{SigmaBecomesT}{
  \hat{T}(\rho) = (\cosh\rho)^{-{d-2 \over d-1}} \left[ \hat{T}_0 + 
    {2(d-2) \over d(d-1)} {\rm H}_0 F_d(\rho) \right] \,.
 }
where $\hat{T}_0$ is the de Sitter temperature at de Sitter time $\rho=0$.  Note that $\hat{n}$ cancels out of the final result \eno{SigmaBecomesT}, as it must.  Using \eqref{TransformHydro} we can map \eqref{SigmaBecomesT} to the energy density on $\mathbf{R}^{d-1,1}$. We find
\begin{equation}
\label{E:Solution}
	\epsilon = \left({\hat{T} \over \tau}\right)^d\,.
\end{equation}
For $d=4$, the solution \eno{E:Solution} reduces to the one found in \cite{Gubser:2010ze}, specified by $\hat{u}_\mu = (-1,0,0,0)$ and
 \eqn{TbDef}{
  \hat{T}(\rho) = {\hat{T}_0 \over (\cosh\rho)^{2/3}} \left[ 1 + 
    {{\rm H}_0 \over 9\hat{T}_0} \sinh^3 \rho \; {}_2F_1\left( {3 \over 2},
      {7 \over 6}, {5 \over 2}, -\sinh^2 \rho \right) \right] \,.
 }
When we wish to investigate numerical properties of this solution, or of generalizations of it, we will usually focus on the parameter choices
 \eqn{SampleParams}{
  \hat{T}_0 = 5.55 \qquad\qquad
  {\rm H}_0 = 0.33 \,.
 }
These parameters, along with
 \eqn{qChoice}{
  q = {1 \over 4.3\,{\rm fm}} \,,
 }
were identified in \cite{Gubser:2010ze} as providing an approximate match to head-on gold-gold collisions at top RHIC energies, $\sqrt{s_{NN}} = 200\,{\rm GeV}$.

\subsection{Properties of the viscous solutions}
 \label{PROPERTIES}
Note that based on \eno{SigmaBecomesT}, since $F_d$ is symmetric in $x$, $\hat{T}(\rho)$ is negative for sufficiently early de Sitter times.  If $\hat{T}_0/H_0 \ll 1$ we can approximate $F_d$ by a power law for very negative values of its argument and find the critical time, $\rho_*$, for which $\hat{T}(\rho_*)=0$. We find
\begin{equation}
\label{E:Approximaterhos}
	\rho_* \sim -\frac{1}{d-2} \ln\left(\frac{d^{d-1}}{2}\right) - \frac{d-1}{d-2} \ln\left(\frac{\hat{T}_0}{{\rm H}_0}\right)\,.
\end{equation}
For $d=4$, and for the parameter choices indicated in \eno{SampleParams}, which according to the analysis of \cite{Gubser:2010ze} are similar to the values of the QGP generated at RHIC, we find that according to \eqref{E:Approximaterhos} $\rho_* \sim -6.09$ which is fairly close to the numerical solution of $\hat{T}(\rho_*)=0$ which is $\rho_* \sim -6.15$. At late de Sitter times the de Sitter temperature approaches a constant value of
\begin{equation}
\label{E:hatTlarger}
	\hat{T} \xrightarrow[\rho \to \infty]{} \frac{2 {\rm H}_0}{d}\,.
\end{equation}

\begin{figure}
\begin{center}
\includegraphics{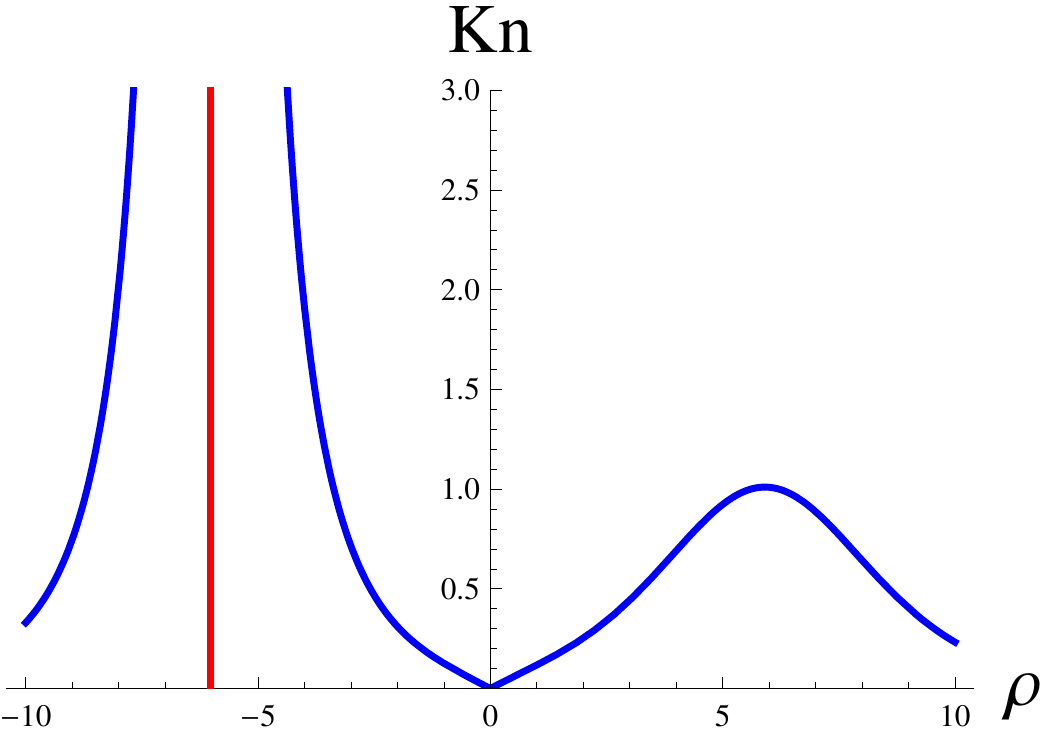}
\caption{\label{F:Knudsen} A plot of the Knudsen number defined in \eqref{E:Knudsen} for the viscous $SO(3)$ symmetric flow given in \eqref{SigmaBecomesT}. The Thick red line marks the location of the expected instability at early times discussed in the text. }
\end{center}
\end{figure}

Negative temperatures is a pathology which signals that the hydrodynamic approximation should not be used at these early times.  In physical terms, the shear is very strong at early times, so the viscous correction terms in the Navier-Stokes equations become comparable to or larger than the pressure terms. Failure of the hydrodynamic approximation can be characterized by the Knudsen number $Kn$ which is given by the ratio of the mean free path to the typical representative scale of the flow.  If the Knudsen number is much larger than one it implies that the continuum assumption necessary for the hydrodynamic approximation has broken down. In a conformal theory the mean free path is proportional to the inverse temperature. For the $SO(d)$-flow in de Sitter space we use
\begin{equation}
\label{E:Knudsen}
	Kn = \frac{1}{\hat{T}} \left| \frac{\hat{T}'}{\hat{T}} \right| \,.
\end{equation}
As expected, the Knudsen number diverges at $\rho=\rho_*$. It vanishes at $\rho=0$, in the $\rho \to \infty$ limit, and at the zero of $\hat{T}'$. A careful analysis of \eqref{SigmaBecomesT} shows us that such a zero exists only when $\rho$ is positive and when the viscosity is large,
\begin{equation}
\label{E:Tpzeroes}
	\frac{\hat{T}_0}{{\rm H}_0} < \frac{\sqrt{\pi} \;\Gamma\left(\frac{d}{2d-2}\right)}{d\; \Gamma\left(\frac{2d-1}{2d-2}\right)}\,.
\end{equation}
For $d=4$ we find that \eqref{E:Tpzeroes} reduces to
 \eqn{ViscosityBoundini}{
  {\hat{T}_0 \over {\rm H}_0} < -{\sqrt\pi \Gamma(-1/3) \over 2\;\Gamma(1/6)}
    \approx 0.65 \,.
 }
A plot of the Knudsen number for the physically interesting values of $d=4$, $\hat{T}_0=5.55$ and ${\rm H}_0 = 0.33$ is given in figure \ref{F:Knudsen}.

\subsection{Second order hydrodynamics}
In \eqref{E:Tmn0and1} we have written down the most general hydrodynamic stress-energy tensor, consistent with conformal symmetries up to first order in gradients. As is well known first order relativistic viscous hydrodynamics suffers from a causality problem (see, for example, \cite{Romatschke:2009im}). That is to say, at large enough momentum superluminal modes may be present in the hydrodynamic solution. This isn't a problem per se since solutions with large momenta are outside the regime of validity of the hydrodynamic approximation. However, these modes do pose a problem when constructing a numerical solution to the Navier Stokes equation \eqref{E:TmnConservation}.

In \cite{Baier:2007ix} the causality problem has been addressed by considering contributions to the stress tensor which contain two derivative terms. It can be shown that in the absence of a charged current and in a conformally flat spacetime the energy momentum tensor takes the form
\begin{equation}
\label{E:Tmn2}
	T^{\mu\nu} = \frac{\epsilon}{d-1} \left( d u^{\mu}u^{\nu} + g^{\mu\nu} \right) -  \eta \sigma^{\mu\nu} +\sum_{i=0}^3 \lambda_i \Sigma^{(i)} + \ldots
\end{equation}
where $\ldots$ denote expressions involving three derivatives of the hydrodynamic variables,
$\lambda_i$ are second order transport coefficients which scale like the temperature squared and the $\Sigma^{(i)}$ are given by
\begin{align}
\begin{split}
\label{E:Sigmas}
	\Sigma^{(0)} &= {}_\langle u^{\lambda}\partial_{\lambda} \sigma_{\mu\nu \rangle} + \frac{1}{d-1}\sigma_{\mu\nu} \partial_{\lambda}u^{\lambda} \,\\
	\Sigma^{(1)} & = \sigma_{\langle \mu \lambda}\sigma^{\lambda}_{\phantom{\lambda}\nu \rangle} \,\quad
	\Sigma^{(2)} = \sigma_{\langle \mu \lambda}\omega^{\lambda}_{\phantom{\lambda}\nu \rangle} \,\quad
	\Sigma^{(3)} = \omega_{\langle \mu \lambda}\omega^{\lambda}_{\phantom{\lambda}\nu \rangle} \,
\end{split}
\end{align}
where
\begin{equation}
	\omega_{\mu\nu} = \frac{1}{2} P_{\mu}^{\lambda}P_{\nu}^{\sigma} \left(\partial_{\lambda}u_{\sigma} - \partial_{\sigma}u_{\lambda}\right)\,.
\end{equation}
See \cite{Baier:2007ix,Bhattacharyya:2008jc} for details.\footnote{When comparing to \cite{Baier:2007ix} it is useful to note that our $\lambda_0$ is related to $\tau_{\pi}$ through $\eta \tau_{\pi} = \lambda_0$.}

We would now like to find an $SO(3)\times SO(1,1)$ symmetric solution to the energy conservation equation \eqref{E:TmnConservation} in the presence of the second order hydrodynamic corrections given in \eqref{E:Tmn2}. Assuming that $\mu=0$ and restricting ourselves to $d=4$, we find that the energy conservation equation reads
\begin{equation}
\label{E:alphaeqn}
	\alpha'+\frac{4}{3} \alpha \tanh \rho-\frac{2}{3} {\rm H}_0 \sqrt{\alpha} (\tanh \rho)^2+\frac{4}{3} {L}_1
   \tanh \rho + \frac{2}{3} {L}_2 \tanh \rho \; (\sech \rho)^2=0
\end{equation}
where $\sqrt{\alpha^2} = \hat{T}^2$ and
\begin{equation}
	\epsilon^{1/2} L_1 = \frac{1}{3}\left( \lambda_0 - \lambda_1 \right)\,, \qquad
	\epsilon^{1/2} L_2 = \frac{1}{3}\left(\lambda_0 + 2\lambda_1 \right)\,.
\end{equation}
The transport coefficients $\lambda_2$ and $\lambda_3$ do not enter into \eqref{E:alphaeqn} since the vorticity $\omega_{\alpha\beta}$ vanishes for the velocity field in \eqref{uChoice}.
Unfortunately, \eqref{E:alphaeqn} is non-linear and difficult to solve. In the case were $L_1=L_2=0$ it reduces to \eqref{SigmaForm} with \eqref{epsilonExample} whose solution is \eqref{SigmaBecomesT}. On the other hand, if we set ${\rm H}_0=0$ but keep $L_1$ and $L_2$ non zero we find that
\begin{equation}
\label{E:Order2sol}
	\alpha = \frac{\hat{T}_0^2}{(\cosh \rho)^{4/3}} - L_1 + \frac{L_2}{(\cosh \rho)^2}\
\end{equation}
is a solution to \eqref{E:alphaeqn}. One can check that when converting this flow back to $\mathbf{R}^{3,1}$ and taking the $q \to \infty$ limit with $\hat{T}_0 q^{-2/3}$ fixed, then the standard solution to Bjorken flow with second order viscous corrections \cite{Baier:2007ix} is recovered.

\begin{figure}
\begin{center}
\includegraphics{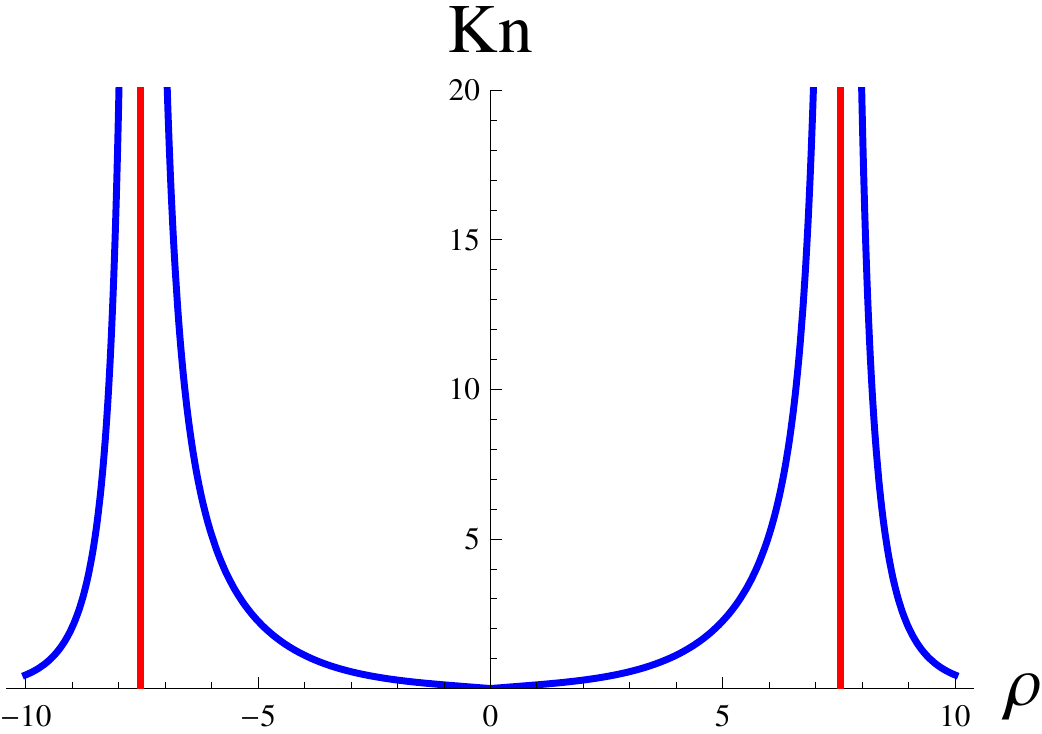}
\caption{\label{F:Knudsen2} A plot of the Knudsen number defined in \eqref{E:Knudsen} for the $SO(3)$ symmetric flow with second order corrections given in \eqref{E:Order2sol} with $L_1$ and $L_2$ as in \eqref{E:SYMVals} and $\hat{T}_0 = 5.5$. The Thick red lines mark the zeros of $\hat{T}$.}
\end{center}
\end{figure}

For $\mathcal{N}=4$ Super Yang-Mills theory with gauge group $SU(N)$ we find that at large $N$ and large t' Hooft coupling \cite{Bhattacharyya:2008jc,Baier:2007ix}\,,
\begin{equation}
\label{E:SYMVals}
	H_0 = \frac{\sqrt{N}}{6^{3/4}\sqrt{\pi}}\qquad
	L_1 = \frac{N}{12\pi\sqrt{6}}\left(1-\ln 2\right) > 0 \qquad 
	L_2 = \frac{N}{12\pi\sqrt{6}}\left(4-\ln 2\right) > 0 \,.
\end{equation}
Since $L_1>0$ one finds that the solution \eqref{E:Order2sol} has vanishing temperature at times $\rho = \pm \rho_*$, $\hat{T}(\pm\rho_{*})=0$, where $\rho_*$ is defined through
\begin{equation}
\label{E:Gotrhostar}
	L_1 (\cosh^{2/3}\rho_{*})^3 - \hat{T}_0 (\cosh^{2/3} \rho_{*}) - L_2 = 0\,.
\end{equation}
The existence of a critical value of $\rho$ for which the temperature vanishes is unusual at first sight, but we have discussed a similar situation in section \ref{PROPERTIES} when we considered first order viscous corrections (${\rm H}_0 \neq 0$, $L_1=L_2=0$) to the $SO(3)$ symmetric flow. When we tried to run the viscous solution \eqref{SigmaBecomesT} backwards in time, we reached a regime where the viscous terms were large relative to the non viscous ones and the hydrodynamic approximation broke down. A similar situation occurs for the second order solution \eqref{E:Order2sol} as can be seen from figure \ref{F:Knudsen2}.
What is strange about \eqref{E:Order2sol} is that due to the time reversal symmetry of \eqref{E:alphaeqn} when setting ${\rm H}_0=0$ there exist two critical values of $\rho$. This may be a peculiarity of fluid flow in de Sitter space, or it may  indicate a breakdown of the hydrodynamic expansion.

\subsection{Angular momentum in $2+1$ dimensions}
\label{MOMENTUM}

The case $d=3$ is special because the symmetry group $SO(2) \times SO(1,1) \times {\bf Z}_2$ doesn't fully determine the velocity profile.  As noted previously, it is consistent with this symmetry to have a non-zero velocity $\hat{u}_\theta$ around the $S^1$ part of $dS_2$.  In the $(\rho,\theta,\eta)$ coordinate system, the general $SO(2) \times SO(1,1) \times {\bf Z}_2$-invariant velocity profile is
 \eqn{lssProfile}{
  \hat{u}_\mu = \left( -{1 \over \sqrt{1-\hat{v}^2}}, {\hat{v} \over \sqrt{1-\hat{v}^2}} 
    \cosh\rho, 0 \right) \,,
 }
where $\hat{v}$ is a function only of $\rho$.  The total momentum around $S^1$ per unit rapidity is
 \eqn{FoundPtheta}{
  \hat{P}_\theta = -\int_0^{2\pi} d\theta \, \cosh\rho \; \hat{T}_{\theta\rho} \,,
 }
and the $SO(2)$ symmetry guarantees that $\hat{P}_\theta$ is constant as a function of de Sitter time $\rho$.  In the inviscid case with zero chemical potential, one can easily show that
 \eqn{PValue}{
  \hat{P}_\theta = 3\pi {\hat{v}\hat\epsilon \over 1-\hat{v}^2} \cosh^2 \rho \,,
 }
and that besides $d\hat{P}_\theta / d\rho = 0$, the only other equation implied by conservation of $T_{\mu\nu}$ is
 \eqn{vDiffEQ}{
  {d\hat{v} \over d\rho} = -{1-\hat{v}^2 \over 2-\hat{v}^2} \hat{v} \tanh\rho \,.
 }
The equation \eno{vDiffEQ} is separable, and the general solution of physical interest is
 \eqn{vlSoln}{
  \hat{v} = \sqrt{\ell (\sqrt{2+\ell^2} - \ell)} \,,
 }
where
 \eqn{ellDef}{
  \ell = \ell_0 \sech\rho
 }
and $\ell_0$ is a constant of integration.  When $\ell_0 \gg 1$, the fluid moves close to the speed of light over a large region of $dS_2 \times {\bf R}$.  We will always pick the positive branch of the square roots in \eno{vlSoln}, thereby obtaining solutions with $\hat{v}>0$; but by choosing the negative branch on the outer square root, one can find solutions with $\hat{v}<0$.  Plugging \eno{vlSoln} into \eno{PValue}, one immediately obtains
 \eqn{EpsilonMomentum}{
  \hat\epsilon = {\hat{P}_\theta \over 3\pi} {\ell^{3/2} \over \ell_0^2}
    {1 - \ell (\sqrt{2+\ell^2} - \ell) \over \sqrt{\sqrt{2+\ell^2} - \ell}} \,.
 }
Taking the $\hat{P}_{\theta} \to 0$ limit together with $\hat{P}_{\theta}/\ell_0^{1/2} \to \hbox{constant}$ one recovers the zero angular momentum solution \eqref{E:Solution}.

\begin{figure}
  \centerline{\includegraphics[width=1.95in]{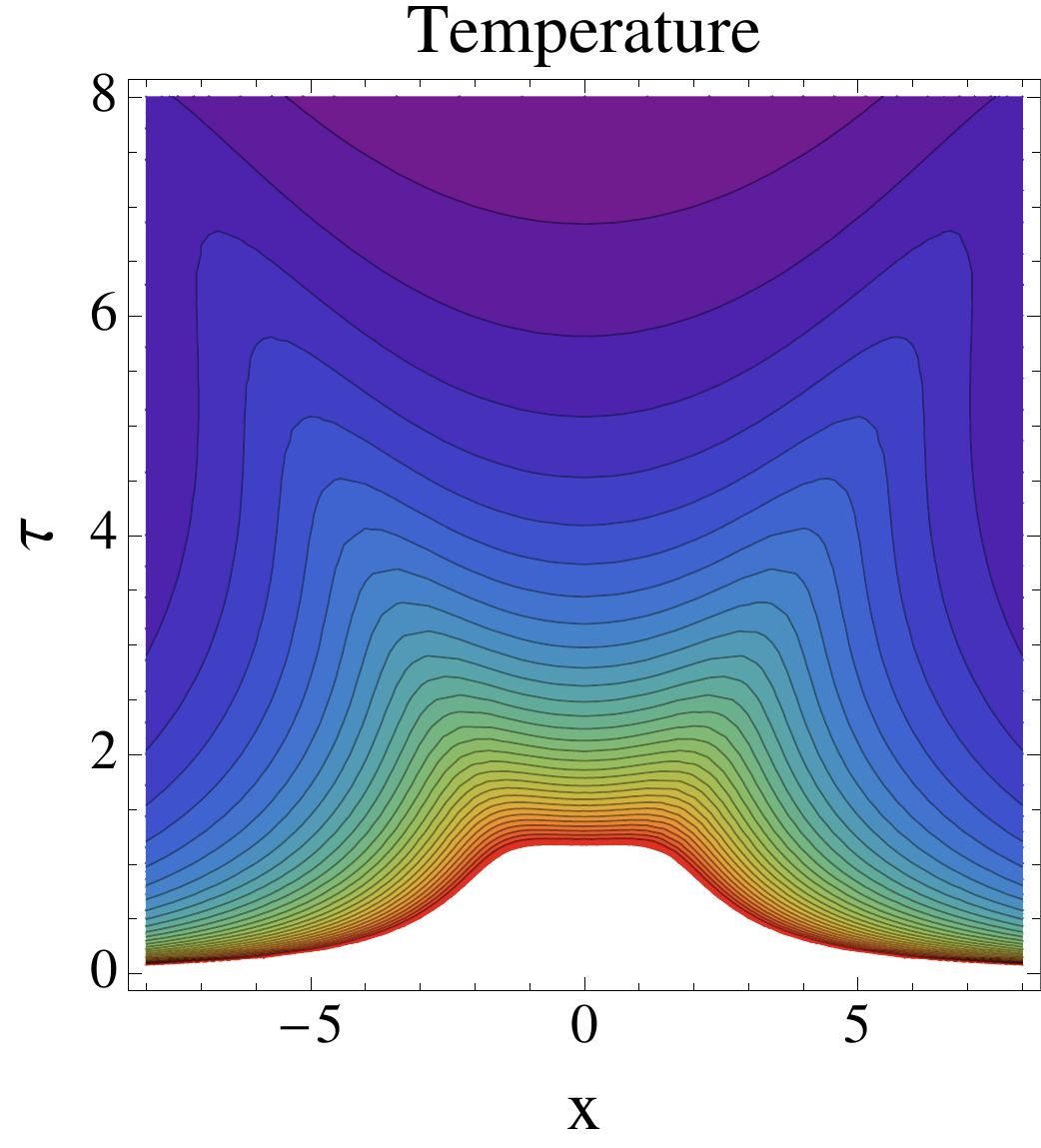}\qquad\includegraphics[width=3.4in]{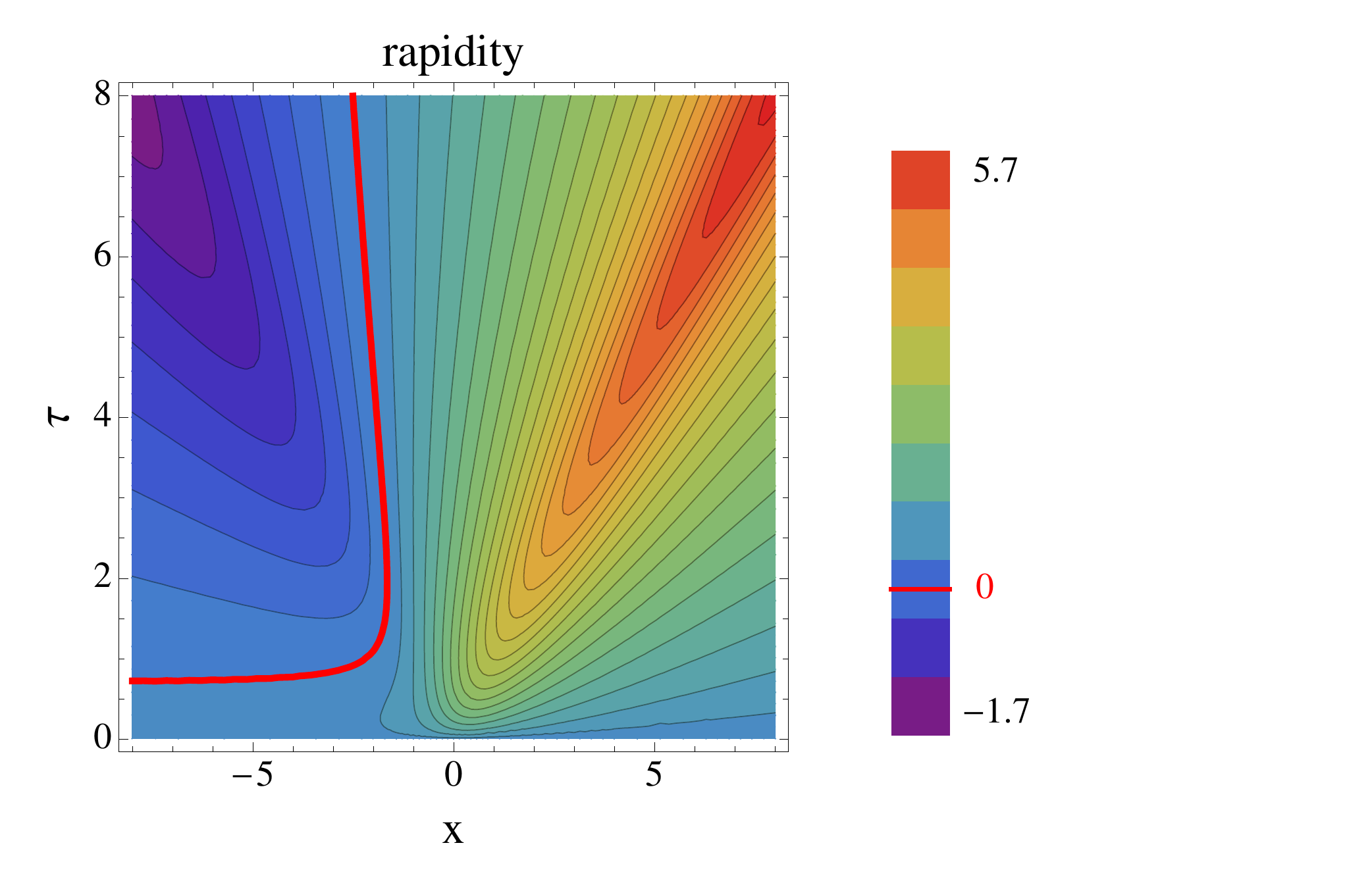}}
  \caption{The temperature and rapidity in the future wedge of ${\bf R}^{2,1}$ for the flow specified in $dS_2 \times {\bf R}$ by the formulas \eno{vlSoln} and \eno{EpsilonMomentum}, with $\ell_0=1$ and units chosen so that $q=1$.}\label{AsymmetricFlow}
 \end{figure}

The formulas \eno{TransformHydro} (with $d=3$) can be used to recover the energy density and velocity profile in the future wedge of ${\bf R}^{2,1}$.  The velocity field in the $(\tau,x,\eta)$ coordinates takes the form
 \eqn{VelocityField}{
  u_\mu = \left( -{1 \over \sqrt{1-v^2}},{v \over \sqrt{1-v^2}},0 \right) \,,
 }
and the flow is completely specified by the two functions $v(\tau,x)$ and $\epsilon(\tau,x)$, whose dependence on $\tau$ and $x$ is entirely algebraic (involving rational functions and square roots only), but too complicated to reproduce here explicitly.  In figure~\ref{AsymmetricFlow} we show an example with $\ell_0 = 1$, plotting the temperature $T = \epsilon^{1/3}$ and the rapidity $\beta = \arctanh v$.  It is interesting to note that the local energy density is symmetric under $x \to -x$ even though the rapidity is strongly asymmetric.

It should be possible to construct flows similar to the one described here in dimensions where the sphere has vector fields which are everywhere non-zero.  It may also be possible to incorporate effects of shear viscosity and/or a non-zero chemical potential.

\section{Embedding in AdS space}
\label{EMBEDDING}

In \cite{Bhattacharyya:2008jc} it was shown that any solution to \eqref{E:TmnConservation} with zero chemical potential corresponds to a black hole configuration in an asymptotically $AdS_5$ geometry, provided the appropriate transport coefficients take on their holographic values.

Explicitly, the construction of \cite{Bhattacharyya:2008jc} is as follows. Suppose that $u_{\mu}$ and $\epsilon$ solve the $d=4$ Navier stokes equation to order $\mathcal{O}(\partial^2)$. By $\mathcal{O}(\partial^2)$ we mean terms which include two derivatives of the velocity field and energy density. For example, $\nabla^2 \epsilon$ is $\mathcal{O}(\partial^2)$, but $u^{\alpha} \nabla_{\alpha} \epsilon$ is $\mathcal{O}(\partial)$. Then, the line element
\begin{equation}
\label{E:GeneralSol}
	ds^2 = -2 u_{\mu} dx^{\mu} (dr + r A_{\nu} dx^{\nu}) + r^{2} g_{\mu\nu} dx^{\mu}dx^{\nu} + \frac{1}{b^4 r^{2}} u_{\mu} u_{\nu} dx^{\mu}dx^{\nu} + b  r^2 F(b r) \sigma_{\mu\nu} dx^{\mu}dx^{\nu} 
\end{equation}
with
\begin{equation}
	A_{\nu} = u^{\lambda} \nabla_{\lambda} u_{\nu} -\frac{1}{3} u_{\nu}\nabla_{\lambda}u^{\lambda}\,,
	\qquad
	b = \frac{1}{\pi T}
\end{equation}
and
\begin{equation}
	F(b r) = \frac{\pi}{4} + \frac{1}{2} \arccoth(b r) - \frac{1}{2} \arctan(b r) + \frac{1}{4}\ln \left(1-(b r)^{-4}\right)
\end{equation}
is a solution to the Einstein equations
\begin{equation}
\label{E:Einstein}
	R_{\mu\nu} - \frac{1}{2} g_{\mu\nu} R - 6 g_{\mu\nu} = 0
\end{equation}
provided terms of order $\mathcal{O}(\partial^2)$ are neglected. An extension of the solution \eqref{E:GeneralSol} which includes second order corrections and a non vanishing chemical potential can be found in \cite{Haack:2008cp,Bhattacharyya:2008mz,Erdmenger:2008rm,Banerjee:2008th}. See \cite{Rangamani:2009xk} for a review.

We emphasize that any solution to the conformally invariant version of the relativistic Navier-Stokes equation \eqref{E:TmnConservation} can be written as a black hole solution in an asymptotically AdS geometry (with a boundary theory on $\mathbf{R}^{3,1}$) by using \eqref{E:GeneralSol}. An explicit and well known example of a solution which has been rewritten as a black hole solution to \eqref{E:Einstein} is Bjorken flow \cite{Janik:2005zt,Janik:2006ft,Heller:2007qt,Heller:2008mb}. Going to Bjorken coordinates,
\begin{equation}
	ds^2 = g_{\mu\nu}dx^{\mu}dx^{\nu} = -d\tau^2 + dx_{\perp}^2 + \tau^2 d\eta^2,
\end{equation}
Bjorken flow is the following solution to \eqref{E:TmnConservation},
\begin{equation}
\label{E:Bjorken}
	u^{\tau} = 1
	\qquad
	T = \left(\frac{\tau_0}{\tau}\right)^{1/3} -\frac{\tilde{H}_0}{2 \tau}
\end{equation}
where $\tau_0$ is an integration constant, $\eta = \tilde{H}_0 \epsilon / T$ (which is a slightly different relation than \eqref{epsilonExample}), and all other components of $u^{\mu}$ vanish. 
The AdS black hole dual of Bjorken flow can be obtained by inserting \eqref{E:Bjorken} into \eqref{E:GeneralSol}, with $1/ b= \pi T$  and setting the shear viscosity to its holographic value, $\tilde{H}_0=1/(3 \pi)$. Unfortunately, it is not so straightforward to check that \eqref{E:GeneralSol} with \eqref{E:Bjorken} is a solution to \eqref{E:Einstein} to order $\mathcal{O}(\partial^2)$. The difficulty lies in keeping track of covariant derivatives of $b$ (proportional to the inverse temperature) and $u^{\mu}$. One way to keep track of these derivative terms is to rescale the $\tau$ coordinate so that 
\begin{equation}
\label{E:tautilde}
	\tau = Q \tilde{\tau}\,.
\end{equation} 
Then \eqref{E:Bjorken} takes the form
\begin{equation}
\label{E:Bjorkentilde}
	u^{\tilde{\tau}} = Q
	\qquad
	\frac{1}{\pi b(\tilde{\tau})}  = \left(\frac{\tilde{\tau}_0}{\tilde{\tau}}\right)^{1/3} -\frac{\tilde{H}_0}{2 Q \tilde{\tau}}
\end{equation}
with $\tilde{\tau}_0$ an integration constant, and after inserting \eqref{E:Bjorkentilde} into \eqref{E:GeneralSol} and using
\begin{equation}
	ds^2 = g_{\mu\nu}dx^{\mu}dx^{\nu} = -Q^2 d\tilde{\tau}^2 + dx_{\perp}^2 + Q^2 \tilde{\tau}^2 d\eta^2,
\end{equation}
one can easily check that \eqref{E:Einstein} are satisfied to order $\mathcal{O}(Q^{-1})$ in a large $Q$ expansion. 

For the $SO(3)$ symmetric solution, which we rewrite as
\begin{equation}
\label{E:SO3sol}
	u{\rho} = -Q \tau \qquad
	\tau T = (\cosh \rho)^{-2/3} \left(T_0 + \frac{1}{9} \frac{\tilde{H}_0}{Q} (\sinh \rho)^3 {}_2F_1\left(\frac{3}{2},\frac{7}{6},\frac{5}{2},-(\sinh \rho)^2 \right) \right)
\end{equation}
on $R^{3,1}$ with line element
\begin{equation}
\label{E:SO3onR31}
	ds^2 = g_{\mu\nu} dx^{\mu} dx^{\nu} = Q^2 \tau^2 \left( -d\rho^2 + (\cosh\rho)^2 \left(d\theta^2 + \sin^2\theta d\phi^2 \right) \right) + Q^2 d\eta^2
\end{equation}
where
\begin{equation}
	\eta = \tilde{H}_0 \epsilon T^{-1} \qquad
	\epsilon  \propto T^4 \qquad
	\tau = (\cosh\rho \cosh\theta_1 - \sinh\rho)^{-1} \,,
\end{equation}
the procedure is similar: We insert \eqref{E:SO3sol} and \eqref{E:SO3onR31} into \eqref{E:GeneralSol} with $b=1/(\pi T)$ and $\tilde{H}_0 = 1/3\pi$ to obtain a solution to the Einstein equations \eqref{E:Einstein} up to order $\mathcal{O}(\partial^2)$ which is equivalent to $\mathcal{O}(Q^{-2})$. The extra factors of $Q$ relative to \eqref{NewFrames} provide a simple book-keeping device, much like \eqref{E:tautilde}, which we use to keep track of gradients of the thermodynamic variables. To derive the solution \eqref{E:SO3sol} with these extra factors of $Q$ one can rescale the spacetime coordinates of the Minkowski metric by a factor of $Q$, similar to \eqref{E:tautilde}, and make appropriate changes in equations \eqref{PoincareToCovering} to \eqref{GlobalToCovering}.

\section{Anisotropies I: setup and equations of motion}
\label{ANISOTROPIES}

The preceding section makes clear that passing from the future wedge of ${\bf R}^{d-1,1}$ to $dS_{d-1} \times {\bf R}$ through a Weyl rescaling is a useful trick for generating exact solutions of conformal relativistic hydrodynamics.  In this section we want to explore its utility in treating perturbations around the $SO(3) \times SO(1,1) \times {\bf Z}_2$-invariant flow with zero chemical potential.  In principle we could include chemical potential and/or consider arbitrary dimensions.  But the case $d=4$ with $\mu=0$ is the most interesting for applications to heavy ion phenomenology, so we will focus entirely on this case.

The calculations in this section are somewhat reminiscent of the analysis of anisotropies in expanding geometries, which form an important part of theoretical cosmology (see for example \cite{Liddle:2000cg} for a pedagogical summary).  There are two important differences.  One is that we are interested ultimately in the contracting half of $dS_3$ (more specifically, the Poincar\'e patch parametrized by $(\tau,\eta,x_\perp,\phi)$), whereas cosmological perturbations are more often considered in expanding geometries.  The other is that gravitational fluctuations are not included; the curved geometry in which we work is nothing more than a fixed Weyl rescaling of the future wedge of Minkowski space.  All we are doing is solving the equations of relativistic hydrodynamics on a fixed, curved background, with perturbations treated (mostly) at linear order.

The analysis of hydrodynamic anisotropies in hydrodynamic flow of the quark-gluon plasma at RHIC has also been recently reconsidered as a mechanism for explaining two-particle correlation measurements: see for example \cite{Alver:2010gr,Alver:2010dn,Petersen:2010cw,Qin:2010pf,Lacey:2010hw,Teaney:2010vd}.  We compare our classification of hydrodynamic perturbations and those of \cite{Teaney:2010vd} in section~\ref{PARAMETRIZE}.

\subsection{Hydrodynamic perturbations}

As usual we do most of our calculations in the $dS_3 \times {\bf R}$ conformal frame, with coordinates $\hat{x}^\mu = (\rho,\theta,\phi,\eta)$, and we use a hat to denote quantities specific to that frame.  Also, we use a subscript $b$ to denote unperturbed background quantities.  Thus, for example, the background temperature of the $SO(3) \times SO(1,1) \times {\bf Z}_2$-invariant solution which we perturb around is denoted $\hat{T}_b$ in this section, whereas in \eno{TbDef} it was denoted $\hat{T}$.

Perturbed fields may be parametrized as follows:
 \eqn{PerturbFlow}{
  \hat{T} = \hat{T}_b (1 + \delta) \qquad
  \hat{u}_\mu = (-1,\nu_i,\nu_\eta) \,,
 }
where $\delta$, $\nu_i$, and $\nu_\eta$ are all to be treated at linear order, and $i$ runs over the two $S^2$ directions. By $\hat{u}_i$ we mean the two $S^2$ components of the four-velocity. Note that linear corrections to the temperature are $\hat{T}_b \delta$.  Corrections to the $\hat{u}_\rho$ component of velocity enter only at quadratic order, so we suppress them.  $\delta$, $\nu_i$, and $\nu_\eta$ can depend on all four coordinates $(\rho,\theta,\phi,\eta)$. 
Our main strategy for computing these perturbations will be to decompose them into scalars and vectors of the $SO(3)$ isometry group of ${dS}_3$.  Explicitly, a basis for hydrodynamic perturbations is
 \eqn{E:Perturbations}{
  \delta(\rho,\theta,\phi,\eta) &= \delta(\rho) S(\theta,\phi) e^{i k_{\eta} \eta}  \cr
  \nu_i(\rho,\theta,\phi,\eta) & = \nu_s(\rho) \partial_i S(\theta,\phi) e^{i k_{\eta} \eta} 
	  + \nu_v(\rho) V_{i}(\theta,\phi) e^{i k_{\eta} \eta}  \cr
  \nu_\eta(\rho,\theta,\phi,\eta) & = \nu_{\eta}(\rho) S(\theta,\phi) e^{i k_{\eta} \eta} \,,
 }
where, by assumption,
\begin{equation}
	\tilde{g}^{ij} \tilde{\nabla}_i V_j = 0 \,,
\end{equation}
and $\tilde{g}^{ij}$ and $\tilde{\nabla}_i$ are the metric and the covariant derivative of the unit $S^2$ parametrized by $(\theta,\phi)$.  Symmetry arguments guarantee that the equations of motion for $\nu_v$ are decoupled from the equations of motion for $\nu_{s}$, $\nu_{\eta}$ and $\delta$. Thus, in solving for the vector modes $\nu_v$, we can set the scalar modes to zero and vice versa.

Practitioners of hydrodynamics will not be surprised to learn that the dynamics of the vector modes $\nu_v$ is purely diffusive.  In section~\ref{VECTOR}, we give a fairly complete account of the vector modes. The scalar modes $(\delta,\nu_s,\nu_\eta)$ are more complicated, exhibiting a combination of diffusive and reactive behavior.  The early time instability mentioned in the introduction is also found in the scalar sector.  We explain the basic equations underlying the scalar modes in section~\ref{SCALAR}, and in section~\ref{SOLUTIONS}  we study some of the solutions to these equations.

In order to get physical insight into what the anisotropies represent, we must transform from the $dS_3 \times {\bf R}$ conformal frame back to more standard coordinates on the future wedge of ${\bf R}^{3,1}$.  This is straightforward.  The flat space form of the perturbed flows is easily found to be
 \eqn{PerturbedFlat}{
  T &= T_b (1 + \delta)  \cr
  u_\tau &= u_\tau^b + \tau {\partial\theta \over \partial\tau} \nu_\theta  \cr
  u_\perp &= u_\perp^b + \tau {\partial\theta \over \partial x_\perp} \nu_\theta  \cr
  u_\phi &= \tau \nu_\phi  \cr
  u_\eta &= \tau \nu_\eta \,,
 }
where by $T_b$ and $u_{\tau}^b$ we mean the unperturbed flow obtained in \eqref{SigmaBecomesT} and \eqref{uChoice} transformed to $\mathbf{R}^{3,1}$ using \eqref{TransformHydro}.

\subsection{Equation of motion for the vector modes}
\label{VECTOR}

A basis of divergenceless vector-valued functions on $S^2$ is provided by the vector spherical harmonics
\cite{Blatt,Hill54,Barrera85} 
 \eqn{PhiDef}{
  { \Phi}_{\ell m}(\theta,\phi) = -\frac{i m Y_{\ell m}(\theta,\phi)}{\sin\theta} \partial_{\theta} +\frac{ \partial_{\theta}Y_{\ell m}(\theta,\phi)}{\sin\theta} \partial_{\phi} \,,
 }
which satisfy
 \eqn{Veqs}{
  \tilde\nabla_i \Phi_{\ell m}^i = 0 \qquad\hbox{and}\qquad
  \tilde\nabla^i \tilde\nabla_i \Phi_{\ell m}^j = -\left[ \ell(\ell+1)-1 \right] 
    \Phi_{\ell m}^j
   \qquad\hbox{for}\qquad \ell = 1,2,\ldots \,,
 }
where indices $i$ are raised and lowered with the metric $\tilde{g}_{ij}$ of the unit $S^2$.  When $\ell=1$, $\Phi^i$ is a Killing vector of $S^2$.

Consider now the excitations \eqref{E:Perturbations} with all the scalar perturbations $(\delta,\nu_s,\nu_\eta)$ set to zero, and with $V_i = \tilde{g}_{ij} \Phi_{\ell m}^j$.  The equation of motion for $\nu_v$ which follows from \eqref{E:TmnConservation} takes the form
\begin{equation}
\label{E:Vmodes}
	\nu_v'(\rho)  = -\Gamma_v(\rho) \nu_v (\rho)
\end{equation}
where
 \eqn{GammaV}{
  \Gamma_v &= {4 \over 9} {T_b \over T_b'} \tanh^2 \rho + 
    {1 \over 3} {{\rm H}_0^2 \over T_b T_b'} \tanh^4 \rho  \cr
    &{}\qquad\qquad{} - 
    {{\rm H}_0 \over 36 T_b'} \left[ 
      -64 + 18 \ell(\ell+1) + 9 k_\eta^2 + (16+9k_\eta^2) \cosh 2\rho \right] 
     \sech^2 \rho \tanh\rho
 }
and primes denote derivatives with respect to $\rho$. 

\subsection{Equation of motion for the scalar modes}
\label{SCALAR}
A basis of scalar functions on $S^2$ is provided by the spherical harmonics $Y_{\ell m}(\theta,\phi)$, which satisfy
 \eqn{BoxY}{
  \tilde\nabla^i \partial_i Y_{\ell m} = -\ell(\ell+1) Y_{\ell m} \,.
 }
Consider the excitations \eno{E:Perturbations} with the vector perturbations $\nu_v$ set to zero, and with $S = Y_{\ell m}$.  If we define
 \eqn{wDef}{
  \vec{w} = \begin{pmatrix} \delta \\ \nu_s \\ \nu_\eta \end{pmatrix} \,,
 }
then instead of a single equation of the form \eqref{E:Vmodes}, energy-momentum conservation gives us the coupled equations
 \eqn{ScalarConserve}{
  \vec{w}' = -{\bf \Gamma}_s \vec{w}
 }
where ${\bf \Gamma}_s$ is a $3 \times 3$ matrix with all entries non-zero: 
 \eqn{Mentries}{
  \Gamma _{s,11} &= \frac{{\rm H}_0 \tanh^2\rho}{3 \hat{T}_b}  \cr
  \Gamma _{s,12} &= \frac{\ell (\ell+1) \sech^2\rho \left[{\rm H}_0 \tanh \rho-\hat{T}_b\right]}{3 \hat{T}_b}  \cr
  \Gamma _{s,13} &= \frac{i k_\eta \left[2 {\rm H}_0 \tanh \rho+\hat{T}_b\right]}{3 \hat{T}_b}  \cr
  \Gamma _{s,21} &= \frac{2 {\rm H}_0 \tanh \rho}{{\rm H}_0 \tanh \rho-2 \hat{T}_b}+1  \cr
  \Gamma _{s,22} &= \frac{{\rm H}_0 \hat{T}_b \left[-4 (3 \ell (\ell+1)-10) \sech^2\rho-9 k_\eta^2-16\right]+6 {\rm H}_0^2 \tanh ^3\rho+8 \hat{T}_b^2 \tanh \rho}{6 \hat{T}_b \left[{\rm H}_0 \tanh \rho-2 \hat{T}_b\right]}  \cr
  \Gamma _{s,23} &= \frac{i {\rm H}_0 k_\eta}{2 {\rm H}_0 \tanh \rho-4 \hat{T}_b}  \cr
  \Gamma _{s,31} &= \frac{i k_\eta \left[3 {\rm H}_0 \tanh \rho+\hat{T}_b\right]}{{\rm H}_0 \tanh \rho+\hat{T}_b}  \cr
  \Gamma _{s,32} &= \frac{i \ell (\ell+1) {\rm H}_0 k_\eta \sech^2\rho}{4 \left[{\rm H}_0 \tanh \rho+\hat{T}_b\right]}  \cr
  \Gamma _{s,33} &= \frac{\hat{T}_b \left[(9 \ell (\ell+1)-4) {\rm H}_0 \sech^2\rho+4 {\rm H}_0 \left[3 k_\eta^2+4\right]-8 \hat{T}_b \tanh \rho\right]+12 {\rm H}_0^2 \tanh ^3\rho}{12 \hat{T}_b \left[{\rm H}_0 \tanh \rho+\hat{T}_b\right]}\,.
 }

Alert readers will notice that there are three possible ways for divergences in components of ${\bf \Gamma}_s$ to arise.  The first route to a divergence in components of ${\bf \Gamma}_s$ is for $\hat{T}_b$ to vanish.  This occurs only at the moment $\rho_*$ when $\hat{T}_b$ first becomes positive.  As discussed in \ref{PROPERTIES} we don't trust hydrodynamics in this region.  The second route to a divergence is for ${\rm H}_0 \tanh\rho - 2\hat{T}_b$ to vanish.  Noting that 
 \eqn{TbpRule}{
  \hat{T}_b'(\rho) = {1 \over 3} \tanh\rho \, ({\rm H}_0 \tanh\rho - 2\hat{T}_b) \,,
 }
we see that the vanishing of ${\rm H}_0 \tanh\rho - 2\hat{T}$ is equivalent to a zero of $\hat{T}_b'$ away from $\rho=0$.  The third route to a divergence is for ${\rm H}_0 \tanh\rho + \hat{T}_b$ to vanish.  It is clear from \eno{GammaV} that $\Gamma_v$ also diverges when $\hat{T}_b=0$ and when $\hat{T}_b'=0$, but not when ${\rm H}_0 \tanh\rho + \hat{T}_b=0$.  In section~\ref{SOLUTIONS} we will explore the physical consequences of these divergences.

\section{Anisotropies II: special limits and stability}
\label{SOLUTIONS}

While linear, equations \eqref{ScalarConserve} and \eqref{E:Vmodes} cannot in general be solved analytically.  One can either resort to numerics, or consider limiting values of the parameters which make the equations tractable.  A key question is whether the $SO(3) \times SO(1,1) \times {\bf Z}_2$-invariant flow is stable against small perturbations.  The answer we will reach is that there are instabilities related to flow in the rapidity direction.  In the viscous case, these instabilities arise in the linearized approximation only at early times---earlier than the times at which hydrodynamics is typically initialized in hydrodynamic simulations of heavy-ion collisions. 

We start in section~\ref{VMODESSolution} with the vector modes, and then continue in section~\ref{SMODESSolution} with the scalar modes, investigating special limits and stability.  We provide in section~\ref{NONLINEAR} an explicit, analytical account of how these perturbations grow large and qualitatively change the flow.

\subsection{Vector modes}
\label{VMODESSolution}

Formally, the solution to \eno{E:Vmodes} is
 \eqn{FormalVector}{
  \nu_v(\rho) = \exp\left\{-\int_{\rho_i}^\rho d\tilde\rho \, \Gamma_v(\tilde\rho)
    \right\} \nu_v(\rho_i) \,,
 }
where $\rho_i$ is an arbitrary initial time.  The integral in \eno{FormalVector} cannot be done explicitly in the viscous case because $\Gamma_v(\rho)$ contains hypergeometric functions.  But in the inviscid case, where
 \eqn{GammaVInviscid}{
	\Gamma_v = -\frac{2}{3}\tanh\rho \,,
 }
one finds immediately that
 \eqn{nuvInviscid}{
  \nu_v(\rho) = (\cosh\rho)^{2/3} \nu_v(0) \,.
 }
The exponential growth of $\nu_v$ at large positive and negative $\rho$ seems to indicate departure from the regime of validity of linearized perturbation theory.  But this is not the case.  The correct validity criterion is that the velocity vector should remain uniformly non-relativistic.  From \eno{PerturbFlow} and \eno{E:Perturbations} we see that the velocity vector is non-relativistic precisely if
 \eqn{NonRelnuV}{
  |\nu_v| \ll \cosh\rho \,,
 }
and this is evidently satisfied by \eno{nuvInviscid}.  So there are no instabilities in the inviscid case, and since \eno{nuvInviscid} has no dependence on $\ell$, the general linearized vector flow can be written down immediately as
 \eqn{GeneralVector}{
  \nu_v(\rho,\theta,\phi,\eta) = (\cosh\rho)^{2/3} \nu_v(0,\theta,\phi,\eta) \,.
 }

Intuitively, the main effect of viscosity should be to cause perturbations to decay, relative to their behavior in the inviscid case, as one moves forward in de Sitter time.  We will consider two limits of $\Gamma_v$ that, for the most part, confirm this intuition.  The first limit is a small viscosity, short wavelength limit:
 \eqn{ShortLimit}{
  {\rm H}_0\ \hbox{small;}\qquad
  \ell\ \hbox{large;}\qquad
  k_\eta\ \hbox{large;}\qquad
  \ell^2 {\rm H}_0\ \hbox{finite;}\qquad
  k_\eta^2 {\rm H}_0\ \hbox{finite;}\qquad
  \rho\ \hbox{finite} \,.
 }
The second one is a late time limit:
 \eqn{LateLimit}{
  {\rm H}_0\ \hbox{finite, non-zero;}\qquad
  \ell\ \hbox{finite;}\qquad
  k_\eta\ \hbox{finite;}\qquad
  \rho \gg 1 \,.
 }
Evidently, these limits are non-overlapping, so they offer complementary insights into the effects of viscosity.

In the small wavelength, low viscosity limit \eno{ShortLimit},
 \eqn{LargeWaveNumber}{
  \Gamma_V = - {2 \over 3} \tanh\rho + 
    {3 {\rm H}_0 (\ell^2 + k_\eta^2 \cosh^2 \rho) \over 4 \hat{T}_0 (\cosh\rho)^{4/3}} + 
      \ldots \,.
 }
The omitted terms in \eno{LargeWaveNumber} are suppressed relative to the ones shown either by positive powers of ${\rm H}_0$, or by negative powers of $\ell$ or $k_\eta$, or both.  Because the second term in \eno{LargeWaveNumber} is everywhere positive, it indeed suppresses perturbations at late de Sitter times relative to the inviscid case.  It so happens that the integral \eno{FormalVector} can be done explicitly using the approximation \eno{LargeWaveNumber}, with the result
 \eqn{FancyNuV}{
  \nu_v(\rho) &= \exp\left\{ {3 {\rm H}_0 \over 4\hat{T}_0} \left[ 
    -{3\ell^2 \over (\cosh\rho)^{1/3}} + (2\ell^2 - k_\eta^2) \,
      {}_2F_1\left( {1 \over 2}, {1 \over 6}; {3 \over 2}; -\sinh^2 \rho \right) \right] 
      \sinh\rho \right\}  \cr
     &\qquad\qquad {}\times (\cosh\rho)^{2/3} \nu_v(0) \,.
 }
When $k_\eta = 0$, the explicit expression \eno{FancyNuV} leads to the following results for the large $\rho$ asymptotics of vector perturbations:
 \eqn{NuVLimit}{
  \nu_v(\rho) &\approx e^{ {3 \ell^2 {\rm H}_0 \over 4 \hat{T}_0} 
    {\sqrt\pi \, \Gamma(-1/3) \over \Gamma(1/6)}} 
    (\cosh\rho)^{2/3} \nu_v(0) \approx 
   e^{-0.97 {\ell^2 {\rm H}_0 \over \hat{T}_0}} (\cosh\rho)^{2/3} \nu_v(0) 
     \qquad\hbox{for $\rho \gg 1$}  \cr
  \nu_v(\rho) &\approx e^{ -{3 \ell^2 {\rm H}_0 \over 4 \hat{T}_0} 
    {\sqrt\pi \, \Gamma(-1/3) \over \Gamma(1/6)}} 
    (\cosh\rho)^{2/3} \nu_v(0) \approx 
   e^{0.97 {\ell^2 {\rm H}_0 \over \hat{T}_0}} (\cosh\rho)^{2/3} \nu_v(0)  
     \qquad\hbox{for $\rho \ll -1$} \,.
 }
It is evident from the form of \eno{LargeWaveNumber} that the suppression of vector modes becomes much stronger when one makes $k_\eta$ non-zero.

In the late time limit \eno{LateLimit},
 \eqn{GammaVLate}{
  \Gamma_v = {3 \over 2^{5/3}} {k_\eta^2 {\rm H}_0 \over 2\hat{T}_0 + 
    {\rm H}_0 {\sqrt\pi \Gamma(-1/3) \over \Gamma(1/6)}} e^{2\rho/3} + {4 \over 3} + 
    {\cal O}(e^{-2\rho/3})\,.
 }
From the first term one sees that the behavior of vector perturbations is entirely stable at late times provided 
 \eqn{ViscosityBound}{
  {\hat{T}_0 \over {\rm H}_0} > -{\sqrt\pi \Gamma(-1/3) \over 2\Gamma(1/6)}
    \approx 0.647 \,.
 }
Comparing \eno{ViscosityBound} to \eno{ViscosityBoundini}, we see that \eno{ViscosityBound} is precisely the criterion for $\hat{T}_b'$ to have no zeros at positive $\rho$.

Now let's consider the case where the condition \eno{ViscosityBound} is violated.  Then for $k_\eta \neq 0$, $\Gamma_v$ becomes exponentially negative at late de Sitter times.  As a result, vector perturbations with $k_\eta \neq 0$ have rapid growth (proportional to the exponential of an exponential of $\rho$) at late de Sitter times.  There is a further peculiarity in $\Gamma_v$ when the condition \eno{ViscosityBound} fails: $\Gamma_v$ has a pole at a finite positive value of $\rho$, call it $\rho_1$, where $\hat{T}'$ vanishes.  Let's consider the effects of such a pole in some general terms, starting by assuming the form
 \eqn{GammaAssume}{
  \Gamma_v(\rho) = {\gamma \over \rho-\rho_1} + \hbox{finite} \,.
 }
From \eno{FormalVector} one immediately concludes that
 \eqn{nuPower}{
  \nu_v(\rho) \approx K_v |\rho-\rho_1|^{-\gamma}
 }
for $\rho$ close to $\rho_1$.  Thus if $\gamma > 0$ there is a divergence in $\nu_v$, signaling the failure of the linearized approximation; while if $\gamma < 0$ there is no divergence, only some failure of differentiability which is presumably smoothed out in the full non-linear theory.  Inspection of $\Gamma_v$ shows that the pole that arises when \eno{ViscosityBound} fails has $\gamma<0$.  Thus there is no problem with the linearized approximation until the strongly negative behavior of the first term of \eno{GammaVLate} drives vector perturbations rapidly out of the linear regime.  We note that for the physically interesting parameters \eno{SampleParams}, the left hand side of \eno{ViscosityBound} is about $26$ times larger than the right hand side.  So fluid flows relevant for heavy-ion phenomenology seem to be safely away from the instabilities that we have described in this paragraph.  If viscous corrections are so large that \eno{ViscosityBound} is violated, one might reasonably ask whether fluid dynamics is applicable at all.

\subsection{Scalar modes}
\label{SMODESSolution}

The scalar sector is considerably more complicated due to the mixing among $\delta$, $\nu_s$, and $\nu_\eta$.  The plan of this section is to investigate solutions to \eno{ScalarConserve} and stability of the scalar sector as far as possible in parallel to our study of vector perturbations in section~\ref{VMODESSolution}.  In particular, we will start with the inviscid case (with one additional simplification), then explain viscous damping in a small wavelength limit, and finally examine stability by studying the late time limit and behavior near singularities of components of ${\bf \Gamma}_s$.

The inviscid case is not analytically tractable in general due to mixing among $\delta$, $\nu_s$, and $\nu_\eta$.  This mixing goes away when $k_\eta=0$, in which case (with ${\rm H}_0$ also set to $0$) we find
 \eqn{GammaSInviscid}{
  {\bf \Gamma}_s = \begin{pmatrix} 0 & -{1 \over 3} \ell (\ell+1) \sech^2 \rho & 0  \\
     1 & -{2 \over 3} \tanh\rho & 0  \\
     0 & 0 & -{2 \over 3} \tanh\rho
    \end{pmatrix} \,.
 }
Evidently, $\nu_\eta$ decouples and obeys the same equation as the vector modes.  So 
 \eqn{nuEtaBig}{
  \nu_\eta(\rho) = (\cosh\rho)^{2/3} \nu_\eta(0) \,.
 }
In the case of the vector modes, we remarked that the exponential growth in $\nu_v(\rho)$ indicated in \eno{nuvInviscid} at late times is not a problem, because the condition for flow to be non-relativistic is $|\nu_v| \ll \cosh\rho$.  The situation is less favorable for $\nu_\eta$: the condition for the flow to be non-relativistic is $|\nu_\eta| \ll 1$, which evidently is violated by the solution \eno{nuEtaBig} for any non-zero $\nu_\eta(0)$.  When $\nu_\eta(\rho)$ gets big, the linear approximation is not justified.  We will soon see that viscosity cures this particular instability, so it can be regarded as an artifact of setting the shear viscosity strictly to zero.

It is straightforward to show starting from \eno{GammaSInviscid} that $\nu_s$ can be eliminated through the equation
 \eqn{uExpress}{
  \nu_s(\rho) = {3 \over \ell (\ell+1)} \cosh^2 \rho {d\delta \over d\rho} \,,
 }
and that $\delta$ obeys the simple second order equation
 \eqn{deltaEq}{
  \left[{d^2 \over d\rho^2} + {4 \over 3} \tanh \rho {d \over d\rho} + 
    {\ell(\ell+1) \over 3} \sech^2 \rho \right] \delta = 0 \,.
 }
This last equation can be solved in terms of associated Legendre functions:
 \eqn{deltaSoln}{
  \delta(\rho) = (\sech\rho)^{2/3} 
    \left[ K_- P_{-{1 \over 2} + {1 \over 6} \sqrt{1+12\ell(\ell+1)}}^{2/3}(-\tanh\rho) + 
           K_+ P_{-{1 \over 2} + {1 \over 6} \sqrt{1+12\ell(\ell+1)}}^{2/3}(\tanh\rho) \right]
           \,,
 }
where $K_\pm$ are integration constants.  If only $K_-$ is non-zero, then $\nu_s$ remains finite as $\rho \to -\infty$ but grows as $e^{2\rho/3}$ as $\rho \to +\infty$.  As with vector modes, this late-time growth is not a concern because the criterion for departure from the non-relativistic regime where linear perturbation theory is good is $|\nu_s| \ll \cosh\rho$.

A good way to understand the overall behavior of the scalar modes is to consider the eigenvalues of ${\bf \Gamma}_s$.  When ${\rm H}_0=k_\eta=0$, these eigenvalues may be read off easily from \eno{GammaSInviscid}:
 \eqn{GammaEigenvalues}{
  \lambda_\pm &= -{1 \over 3} \tanh\rho \pm
    {1 \over 3} \sech\rho \, \sqrt{\sinh^2 \rho - 3 \ell(\ell+1)}  \cr
  \lambda_\eta &= -{2 \over 3} \tanh\rho \,.
 }
Clearly, $\lambda_\eta$ is the eigenvalue related to the diffusive mode $v_\eta$, while $\lambda_\pm$ relate to the $(\delta,\nu_s)$ modes¶.  These modes are oscillatory when $\lambda_\pm$ are complex, which is to say when $|\rho|$ is not too large.  At large $|\rho|$, the $(\delta,\eta_s)$ modes are non-oscillatory.

Now let's consider what happens when shear viscosity is added.  As in section~\ref{VMODESSolution}, we find it useful to consider two complementary limits.  The first one is a small viscosity, short wavelength limit:
 \eqn{ShortLimitScalar}{
  {\rm H}_0\ \hbox{small;}\qquad
  \ell\ \hbox{large;}\qquad
  k_\eta=0;\qquad
  \ell^2 {\rm H}_0\ \hbox{finite;}\qquad
  \rho\ \hbox{finite} \,.
 }
The only reason to require $k_\eta=0$ as part of the limit \eno{ShortLimitScalar} is that it substantially simplifies the equations.  The second limit we consider is
 \eqn{LateLimitScalar}{
  {\rm H}_0\ \hbox{finite, non-zero;}\qquad
  \ell\ \hbox{finite;}\qquad
  k_\eta\ \hbox{finite;}\qquad
  \rho \gg 1 \,,
 }
which is identical to \eno{LateLimit}.

In the small wavelength, low viscosity limit \eno{ShortLimitScalar},
 \eqn{GammaSMod}{
  {\bf \Gamma}_s &= \begin{pmatrix} 0 & -{1 \over 3} \ell (\ell+1) \sech^2 \rho & 0  \\
     1 & -{2 \over 3} \tanh\rho & 0  \\
     0 & 0 & -{2 \over 3} \tanh\rho
    \end{pmatrix}  \cr
   &\qquad{} + {{\rm H}_0 \over \hat{T}_0 (\cosh\rho)^{4/3}}
    \begin{pmatrix} 0 & {1 \over 3} \ell(\ell+1) \tanh\rho & 0 \\
      0 & \ell^2 & 0 \\
      0 & 0 & {3 \over 4} \ell^2
    \end{pmatrix} + \ldots \,,
 }
where, as in \eno{LargeWaveNumber}, the omitted terms are suppressed relative to the ones shown either by positive powers of ${\rm H}_0$, or by negative powers of $\ell$, or both.  Clearly, the matrix in the second line of \eno{GammaSMod} shifts $\lambda_\eta$ by $+{3\ell^2 {\rm H}_0 \over 4 \hat{T}_0 (\cosh\rho)^{4/3}}$, just as in \eno{LargeWaveNumber}.  So once again, $\nu_\eta$ formally behaves just like the vector modes.  The $(\delta,\nu_s)$ modes are more complicated.  Numerical investigation shows that the shift of $\lambda_\pm$ induced by viscous corrections in the limit \eno{LateLimitScalar} is not uniformly positive when $\lambda_\pm$ are purely real; but it is usually positive.  In particular, one can show by explicit calculation that during the oscillatory phase, where $\lambda_\pm$ are complex, the effect of viscous corrections on $\Re\lambda_\pm$ is to shift it by $+{\ell^2 {\rm H}_0 \over 2 \hat{T}_0 (\cosh\rho)^{4/3}}$.  All this is generally in line with the intuitive expectation that the viscosity damps out all perturbations, relative to their inviscid evolution, as $\rho$ becomes more positive.

Now let us turn to the late time limit \eno{LateLimitScalar}, where one finds
 \eqn{GammaSLate}{
  {\bf \Gamma}_s = {\rm H}_1 e^{2\rho/3}
    \begin{pmatrix} 0 & 0 & 0 \\  -{4 \over 3} & k_\eta^2 & -{i k_\eta \over 3} \\
       0 & 0 & 0 \end{pmatrix} + 
    \begin{pmatrix} {2 \over 3} & 0 & {5i \over 3} k_\eta \\  1 & {4 \over 3} & 0 \\
      {7i \over 3} k_\eta & 0 & 2 + {2 \over 3} k_\eta^2 \end{pmatrix} + {\cal O}(e^{-2\rho/3})
     \,,
 }
where
 \eqn{HoneDef}{
  {\rm H}_1 \equiv {3 \over 2^{5/3}} {{\rm H}_0 \over 2 \hat{T}_0 + {\rm H}_0
     {\sqrt\pi \Gamma(-1/3) \over \Gamma(1/6)}} \,.
 }
Note that there is no $\ell$ dependence in ${\bf \Gamma}_s$ to the order shown in \eno{GammaSLate}.  The eigenvalues of ${\bf \Gamma}_s$ as given by \eno{GammaSLate} are easy to find explicitly:
 \eqn{ThreeLambdas}{
  \lambda_1 &= {1 \over 3} \left( 4 + k_\eta^2 - \sqrt{4 - 31 k_\eta^2 + k_\eta^4} \right)  \cr
  \lambda_2 &= {1 \over 3} \left( 4 + k_\eta^2 + \sqrt{4 - 31 k_\eta^2 + k_\eta^4} \right)  \cr
  \lambda_3 &= {\rm H}_1 k_\eta^2 e^{2\rho/3} + {4 \over 3} \,. 
 }
It is clear from the first term in $\lambda_3$ that late stability for $k_\eta \neq 0$ holds only ${\rm H}_1 > 0$, which is the same condition as the bound \eno{ViscosityBound}.  The pathologies that arise when this bound is violated, together with the observation that violation of the bound is far from the regime of interesting parameters in heavy ion collisions, dissuade us from considering ${\rm H}_1 < 0$ here.  Instead we will focus on the case ${\rm H}_1>0$: then all three eigenvalues $\lambda_i$ have positive real parts for arbitrary real $k_\eta$.  This is generally indicative of stability.  Numerical exploration confirms that all three quantities $(\delta,\nu_s,\nu_\eta)$ are driven to small values at late times $\rho$.  Thus, shear viscosity cures the late de Sitter time instability in $\nu_\eta$.

The only remaining route to instability is for some component of ${\bf\Gamma}_s$ to diverge at finite $\rho$, in the region where $\hat{T}_b > 0$. This indeed happens, in two different ways.  The first way is that if ${\rm H}_1 < 0$, three components of ${\bf \Gamma}_s$ diverge at some $\rho_1>0$.  As stated earlier, this instability is associated with zeros of $\hat{T}'_b$, and we will not explore it further.  The last type of divergence arises when ${\rm H}_0 \tanh\rho + \hat{T}_b$ vanishes.  This occurs at some de Sitter time $\rho_2$ which is negative, but less negative than the time $\rho_*$ where $\hat{T}_b$ first becomes positive.  Using the identity \eno{TbpRule}, one can show that the Knudsen number at $\rho=\rho_2$, as defined in \eno{E:Knudsen}, is $1/{\rm H}_0$.  Keeping in mind that ${\rm H}_0 = 0.33$ is in the ballpark of realistic numbers for a heavy ion collision at top RHIC energies, we see that fluid dynamics is not particularly reliable at $\rho=\rho_2$.  Indeed, $\rho=\rho_2$ corresponds to proper time $\tau \approx 0.05\,{\rm fm}/c$ at $x_\perp = 0$ (see section~\ref{EXAMPLES} for a fuller discussion) which is a substantially earlier time than is usually considered reasonable for hydrodynamic simulations to start.  Despite these observations, we wish to understand more fully what happens to the hydrodynamic description at $\rho=\rho_2$.

To simplify our investigation of the divergence at $\rho=\rho_2$, let's start by assuming $k_\eta=0$.  Then $\nu_\eta$ decouples from $\delta$ and $\nu_s$, and the only component of ${\bf \Gamma}_s$ that diverges at $\rho=\rho_2$ is $\Gamma_{s,33}$.  Thus we only need to consider the evolution of $\nu_\eta$.  Near $\rho=\rho_2$, one can show from \eno{Mentries} and \eno{TbpRule} that
 \eqn{GammasResidue}{
  \Gamma_{s,33} = {\gamma \over \rho-\rho_2} + {\rm finite} \qquad\hbox{with}\qquad
   \gamma = 1 + {3 \over 4} \ell(\ell+1) \sech^2 \rho_2 \,.
 }
Evidently, $\gamma>0$ for all $\ell$.  This means that $\nu_\eta$ diverges as $|\rho-\rho_2|^{-\gamma}$.  The divergence is stronger for larger $\ell$.  Numerical exploration indicates that when $k_\eta \neq 0$, the divergence is still present; in fact it spreads to all three quantities $(\delta,\nu_s,\nu_\eta)$.  To say more about this divergence, we need to go beyond the linearized approximation.  We take some steps in this direction in section~\ref{NONLINEAR}.

\subsection{Non-linear treatment of rapidity perturbations}
\label{NONLINEAR}

We learned in section \eno{SMODESSolution} that the perturbation that most often exhibits instabilities is $\nu_\eta$.  We would like to go beyond the linearized treatment of this perturbation.  To do so, we assume $\ell=0$ and $k_\eta=0$: that is, $SO(3) \times SO(1,1)$ symmetry is preserved.  But we allow the ${\bf Z}_2$ symmetry that sends $\eta \to -\eta$ to be broken.  The most general hydrodynamic ansatz consistent with these symmetries is
 \eqn{BoostingFlow}{
  \hat{u}_\rho = -\cosh\chi(\rho) \,, \qquad
  \hat{u}_\theta = \hat{u}_\phi = 0 \,, \qquad
  \hat{u}_\eta = \sinh\chi(\rho) \,.
 }
Temperature $\hat{T}$ (or, equivalently, energy density $\hat\epsilon$) must also be a function only of $\rho$.

With the ansatz \eno{BoostingFlow}, the Euler equations read
 \eqn{InviscidBoosting}{
  {d\chi \over d\rho} &= {\sinh 2\chi \over 2 + \cosh 2\chi} \tanh\rho  \cr
  {d\log \hat{T} \over d\rho} &= -{2\cosh^2 \chi \over 2 + \cosh 2\chi} \tanh\rho
 }
These two equations are readily solved because the first of them is separable.  The general solution can be given in implicit form as
 \eqn{ImplicitBoosting}{
  \cosh^2 \rho &= K_1 \sinh^2 \chi \tanh\chi  \cr
  \hat{T} &= K_2 \csch\chi
 }
where $K_1$ and $K_2$ are constants of integration, which must have the same sign in order for $\hat{T}$ to be positive.  We assume that $K_1$ and $K_2$ are positive, which is the same as choosing $\chi$ to be positive.  Note that at late times,
 \eqn{LateInviscid}{
  \chi &\approx \rho - {1 \over 2} \log K_1  \cr
  \hat{T} &\approx 2 \sqrt{K_1} K_2 \, e^{-\rho} \,.
 }
This is a non-linear version, for $\ell=0$, of the instability of the strict inviscid limit which we noted following \eno{nuEtaBig}.

Now let us pass to the viscous case.  Plugging the ansatz \eno{BoostingFlow} into the viscous Navier-Stokes equations (with vanishing chemical potential) leads to the following two equations:
 \eqn{PetaConstant}{
  {dP_\eta \over d\rho} &= 0  \cr
  {d \over d\rho} \left( \cosh\rho \; \hat{T} \right) &= 
    -{P_\eta \over 16\pi {\rm H}_0 \hat{T}^2} \sech\rho \csch\chi \sech^2 \chi + 
    {3P_\eta^2 \over 256 \pi^2 {\rm H}_0 \hat{T}^6} \sech^3 \rho \csch^2 \chi \sech^3 \chi \,,
 }
where 
 \eqn{PetaViscous}{
  P_\eta &\equiv 4\pi \cosh^2 \rho \; T^\rho{}_\eta  \cr
   &= {16\pi \over 3} \cosh^2 \rho \cosh\chi \sinh\chi \; \hat{T}^4 \left[
       1 + {H_0 \over \hat{T}} \left( \tanh\rho \cosh\chi - 
         {d\chi \over d\rho} \sinh\chi \right) \right] \,
 }
is the momentum in the $\eta$ direction per unit rapidity.
We were unable to find the general solution to the equations \eno{PetaConstant}.  However, an interesting special case is $P_\eta = 0$, corresponding to flows that have no net momentum in the $\eta$ direction.  One way to achieve $P_\eta = 0$ is to set $\chi(\rho) = 0$ everywhere.  Then one is led back to the $SO(3) \times SO(1,1) \times {\bf Z}_2$-invariant solution \eno{TbDef}.  Let us instead assume $\chi \neq 0$, but nevertheless $P_\eta = 0$.  Then the second equation in \eno{PetaConstant} can be used to conclude that $\hat{T} \propto \sech\rho$.  Also, the quantity in square brackets in \eno{PetaViscous} must vanish.  Combining these constraints, one sees that the general solution to \eno{PetaConstant} with $\chi \neq 0$ and $P_\eta = 0$ is 
 \eqn{GeneralBoosting}{
  \cosh\chi = K_3 \cosh(\rho-\rho_0) \qquad\qquad
  \hat{T} = -{\rm H}_0 K_3 \sinh\rho_0 \sech\rho \,,
 }
where $K_3$ and $\rho_0$ are constants of integration. If $K_3=1$, and we assume that $\chi$ is increasing as a function of $\rho$, then we have the simpler form
 \eqn{BoostingSolution}{
  \chi(\rho) = \rho - \rho_0 \qquad
  \hat{T}(\rho) = -{\rm H}_0 \sinh \rho_0 \sech\rho \,.
 }
It is interesting to note some properties of the flows \eno{GeneralBoosting} in other frames.  In flat space,
 \eqn{FlatSpaceSoln}{
  T &= {-2 q {\rm H}_0 K_3 \sinh\rho_0 \over \sqrt{1 + 2q^2 (\tau^2 + x_\perp^2) + 
    q^4 (\tau^2 - x_\perp^2)^2}}  \cr
  v_\perp &\equiv -{u_\perp \over u_\tau} = {2q^2 \tau x_\perp \over 1 + 
    q^2 \tau^2 + q^2 x_\perp^2} \,.
 }
The expression for $v_\perp$ is unchanged from the $SO(3) \times SO(1,1) \times {\bf Z}_2$-invariant flow: it is simply the expression of the condition $\hat{u}_\theta = 0$ in flat space coordinates.  An expression for $v_\eta \equiv -{u_\eta \over \tau u_\tau}$  can also be given, but it is long and unenlightening.  It is worth noting, however, that for the special solutions \eno{BoostingSolution}, the range of $v_\eta$ is the whole interval $(-1,1)$ allowed by causality.  

In the $AdS_2 \times S^2$ conformal frame described in \eno{AdSFrame}, the solutions \eno{GeneralBoosting} satisfy
 \eqn{AdSFrameSolnProps}{
  \mathring{T} = \cosh\rho \; \hat{T} = \hat{T}_0 \qquad\qquad
  \mathring\nabla^\mu \mathring{u}_\mu = {\hat{T}_0 \over {\rm H}_0} \,,
 }
where $\hat{T}_0 = -{\rm H}_0 K_3 \sinh\rho_0$, and we use rings, as on $\mathring{T}$, to denote quantities in the $AdS_2 \times S^2$ frame.  The simplicity of the formulas \eno{AdSFrameSolnProps} suggests that further study of solutions with non-zero $\chi$ would be easiest in the $AdS_2 \times S^2$ frame.

\begin{figure}
  \centerline{\includegraphics[width=7in]{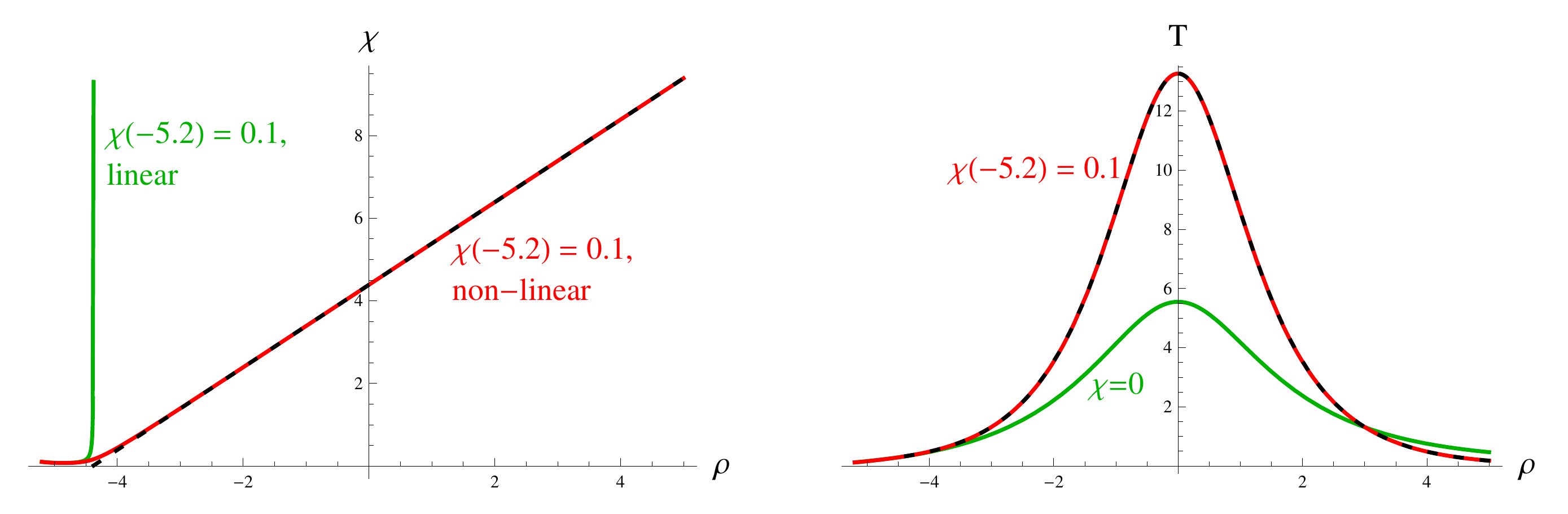}}
  \caption{Left: The rapidity $\chi$ as a function of de Sitter time $\rho$, both in the linear approximation (green) and in the full non-linear theory (red).  Right: The temperature of the $SO(3) \times SO(1,1) \times {\bf Z}_2$-invariant flow (green) and of a flow with non-zero $\chi$ (red). The viscosity parameter is ${\rm H}_0 = 0.33$, as in \eno{SampleParams}.  The specific solutions plotted are based on choosing initial conditions at $\rho_i = -5.2$ such that $\hat{T}$ matches the flow \eno{TbDef} with parameters chosen as in \eno{SampleParams}, and such that $\chi(-5.2) = 0.1$, with $d\chi/d\rho$ chosen in the non-linear treatment to match its value in the linearized treatment.  The dashed black curves show the analytical solution \eno{BoostingSolution} with $\rho_0$ chosen so that $\hat{T}(0)$ matches exactly between the numerical and analytical solutions.}\label{NonlinearFlow}
 \end{figure}

Now let's compare the non-linear flows \eno{BoostingSolution} to the linearized analysis of rapidity fluctuations.  Numerical exploration indicates that if one starts at $\rho < \rho_2$ with a slight perturbation of the $SO(3) \times SO(1,1) \times {\bf Z}_2$-symmetric flow \eno{TbDef} toward non-zero $v_\eta$, with $\ell=k_\eta=0$, the resulting flow quickly converges approximately to \eno{BoostingSolution} with $\rho_0 \approx \rho_2$.  See for example figure~\ref{NonlinearFlow}.  This convergence is not surprising: the initial conditions clearly specify a flow with small $P_\eta$, and we know from the linearized analysis that $\chi$ cannot stay small.  So the general solution \eno{GeneralBoosting} is a natural candidate to describe the evolution.  Picking $\rho_0=\rho_2$ is natural because this is the time when the rapidity starts departing from the linear regime.  It is less clear to us why the choice $K_1=1$ works as well as it does in describing the flow for $\rho > \rho_2$.

Three additional points are worth noting:
 \begin{itemize}
  \item The flows \eno{GeneralBoosting} do not in general have a small Knudsen number.  Using the definition \eno{E:Knudsen}, we have immediately $Kn = {|\sinh\rho| \over \hat{T}(0)}$, which is small in some neighborhood of $\rho=0$ and large elsewhere.  It is perhaps fairer to define $Kn = {|\chi'(\rho)| \over \hat{T}(\rho)}$, and, for the special case $K_1=1$, this leads to the expression $Kn = {\cosh\rho \over \hat{T}(0)}$, which has similar qualitative features to the previous expression provided $\hat{T}_0 \gg 1$.
  \item Solutions to the non-linear equations preserving $SO(3)$ but not $SO(1,1)$ can probably be generated starting from perturbations with non-zero $k_\eta$.  Such solutions are significantly harder to study because the differential equations to be solved have two independent variables, $\tau$ and $\eta$.  Presumably the non-linear evolution leads to rapidity gaps.  Such gaps are not observed in heavy-ion data as far as we are aware.
  \item According to \eno{GammasResidue}, there are modes with $\ell \neq 0$ which show similar instabilities to the one in $v_\eta$ at $k_\eta=\ell=0$ which we have analyzed in this section at the non-linear level.  It therefore seems likely that the full $3+1$-dimensional dynamics is chaotic, with many sound mode instabilities, all related to the one studied here, but with non-zero $\ell$ and $k_\eta$, developing simultaneously.
 \end{itemize}

Let us end this section with a discussion of singularities.  The singularity at $\rho=\rho_2$ noted in the linear theory at the end of section~\ref{SMODESSolution} is smoothed out in the non-linear theory.  However, the non-linear theory has singularities of a new sort: If $K_3 < 1$ in \eno{GeneralBoosting}, then $\chi \to 0$ at two finite values of $\rho$, call them $\rho_\pm$, with $\rho_- < \rho_+$.  For $\rho \in (\rho_-,\rho_+)$, $\chi(\rho)$ is undefined, and as one approaches $\rho_-$ from below, $\chi \sim \sqrt{\rho_- - \rho}$.  Thus the flow becomes non-differentiable at $\rho = \rho_-$.  There is a similar singularity at $\rho=\rho_+$.

\section{Applications to heavy-ion physics}
\label{APPLICATIONS}

As already noted in \cite{Gubser:2010ze} and around equations \eno{SampleParams} and \eno{qChoice}, the $SO(3) \times SO(1,1) \times {\bf Z}_2$-symmetric flow with $\hat{T}_0 = 5.55$, ${\rm H}_0 = 0.33$, and $q = 1/(4.3\,{\rm fm})$ is in the correct ballpark to describe the hydrodynamic evolution of head-on gold-gold collisions at top RHIC energies, $\sqrt{s}_{NN} = 200\,{\rm GeV}$.  This description is obviously imperfect: neither the initial conditions nor the hydrodynamic behavior are truly conformally symmetric in real heavy-ion collisions.  Nevertheless, the $SO(3) \times SO(1,1) \times {\bf Z}_2$-symmetric flow is useful because it correctly captures aspects of real heavy-ion collisions in a formalism which is to a large extent analytically tractable.  The purpose of this section is to provide some translations between aspects of heavy-ion dynamics and the mathematical formalism based on conformal symmetry.  In particular, we will map out in section~\ref{ISOTHERMS} the approximate locations of thermalization and hadronization in the $dS_3 \times {\bf R}$ conformal frame; we will provide in section~\ref{EXAMPLES} several examples of perturbed solutions exhibiting elliptic and triangular flow; and we will explain in section~\ref{PARAMETRIZE} how the natural treatment of initial conditions based on conformal symmetry compares to the cumulant expansion of \cite{Teaney:2010vd}.

\subsection{The history of a heavy ion collision in the de Sitter conformal frame}
\label{ISOTHERMS}

\begin{figure}
  \centerline{\includegraphics[width=4in]{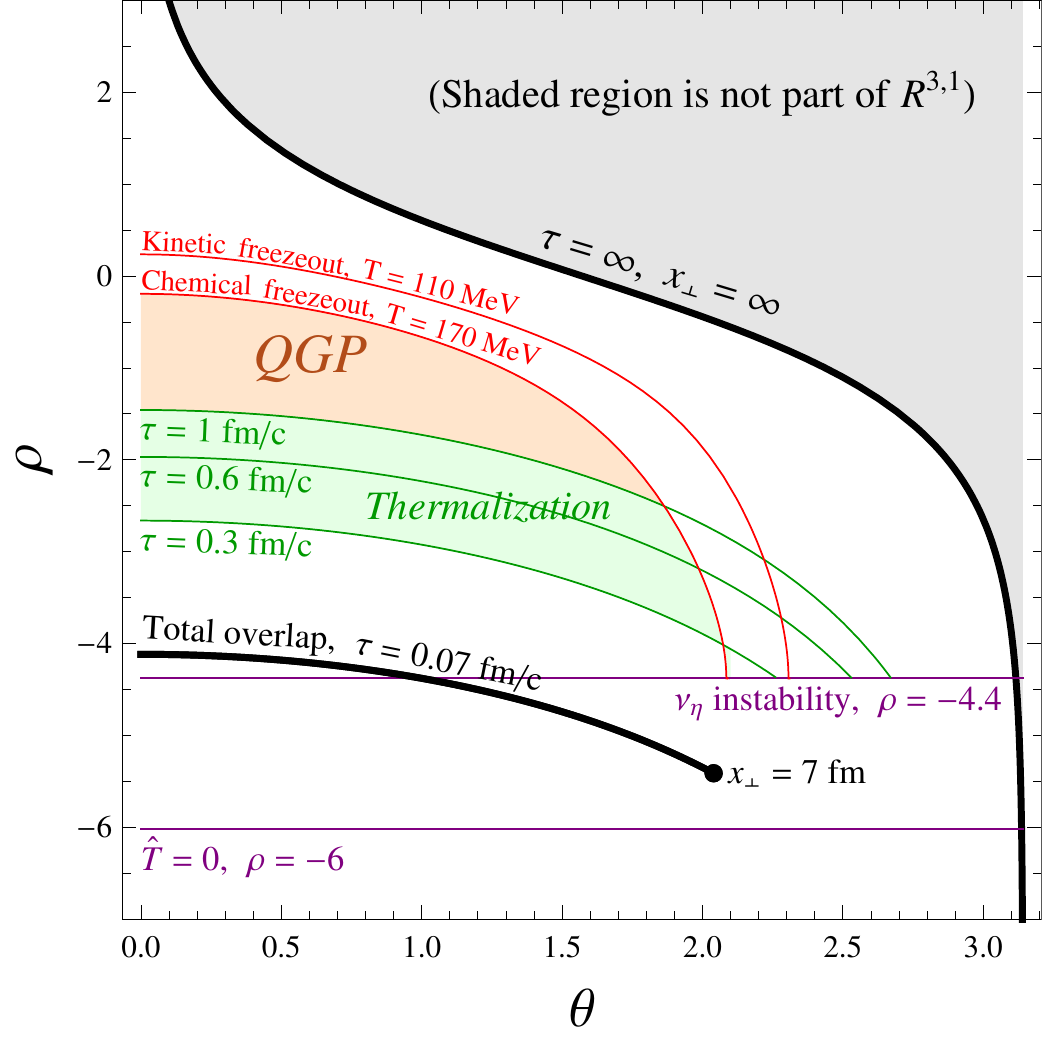}}
  \caption{(Color online.)  An approximate history of a head-on heavy ion collision as seen in the $dS_3 \times {\bf R}$ conformal frame.  Fluid dynamics can be used in the orange region labeled ``QGP.''  If thermalization occurs earlier than $\tau = 1\,{\rm fm}/c$, then fluid dynamics is also valid across part or all of the green region labeled ``Thermalization.''  The thick curve labeled ``total overlap'' shows a characteristic time $\tau = 0.07\,{\rm fm}/c$ for total overlap of the colliding nuclei.  This curve ends at $x_\perp = 7\,{\rm fm}$ because that is approximately the full transverse radius of a gold nucleus.  The line labeled ``$\nu_\eta$ instability'' shows where the instability described in sections~\ref{SMODESSolution} and \ref{NONLINEAR} occurs.  The parameters \eno{SampleParams} and \eno{qChoice} were used in drawing all curves in the figure.  We also used the $SO(3) \times SO(1,1) \times {\bf Z}_2$-invariant solution \eno{TbDef} in order to calculate the position of the curves depicting freezeout, and also of the lines showing where $\hat{T}$ vanishes and where the $\nu_\eta$ instability occurs.}\label{TimeSequence}
 \end{figure}

In this section we aim to show where in $dS_3 \times {\bf R}$ thermalization and hadronization occur.  We will continue to focus on head-on collisions at top RHIC energies (that is, $\sqrt{s_{NN}} = 200\,{\rm GeV}$ for gold-gold collisions).  Our results are presented in figure~\ref{TimeSequence}, which shows how the anatomy of the collision in the $(\tau,x_\perp)$ plane gets repackaged in the coordinates $(\rho,\theta)$.

First let's explain how we plotted the region where thermalization occurs in the $(\rho,\theta)$ coordinates.  Typical choices of thermalization time in hydrodynamic simulations include Bjorken $\tau$ between $0.6\,{\rm fm}/c$ and $1\,{\rm fm}/c$, though values as small as $0.3\,{\rm fm}/c$ are probably acceptable.  From \eno{RhoThetaCoords} one can derive the formulas
 \eqn{tauxpForm}{
  q \tau = {\sech\rho \over \cos\theta - \tanh\rho} \qquad
  q x_\perp = {\sin\theta \over \cos\theta - \tanh\rho} \,.
 }
Using these formulas one can find the locus of points $(\rho,\theta)$ that corresponds to a given $\tau$ (or to a given $x_\perp$).  In figure~\ref{TimeSequence} we show the curves $\tau=0.3\,{\rm fm}/c$, $0.6\,{\rm fm}/c$, and $1\,{\rm fm}/c$.

Hadronization occurs not on a specific time-slice, but through processes at characteristic temperatures.  Chemical freeze-out, determined by thermal fits to hadron yields, occurs at $T_{\rm chem} \approx 170\,{\rm MeV}$ (see for example \cite{BraunMunzinger:2003zd}), whereas kinetic freeze-out, determined by thermal fits to the momentum dependence of particle yields of a fixed species, occurs at $T_{\rm kin} \approx 110\,{\rm MeV}$ (see for example \cite{Heinz:2004qz}).  These temperatures are low enough that deviations from the conformal equation of state, $\epsilon = g_* T^4$ are large, and conformal methods are probably no longer useful.  Nevertheless, in order to get an approximate idea of where in $dS_3 \times {\bf R}$ the validity of fluid dynamics ends, we will use the relation
 \eqn{epsilonTranslate}{
  \epsilon = \left( {\hat{T} \over \tau} \right)^4 = f_* T^4 \,,
 }
with $f_* = 11$, in order to find the locus of points $(\rho,\theta)$ where a given flow attains a definite Minkowski space temperature.  The value $f_* \approx 11$ can be extracted from lattice data \cite{Karsch:2001cy,Gubser:2008pc} for $1.2 T_c < T < 2 T_c$.  In figure~\ref{TimeSequence} we show the curves $T = 110\,{\rm MeV}$ and $T = 170\,{\rm MeV}$.

\subsection{Examples of perturbed flows}
\label{EXAMPLES}

In sections \eqref{VMODESSolution} and \eqref{SMODESSolution} we constructed analytic solutions for the linearized perturbations around the inviscid $SO(3)$-invariant flow. When considering the $SO(3)$-invariant flow as a toy model for the thermodynamic phase of heavy ion collisions, the $\ell=2$ perturbations may be thought of as elliptic flow and the $\ell=3$ perturbations as triangular flow. 

\begin{figure}
  \centerline{\includegraphics[width=7in]{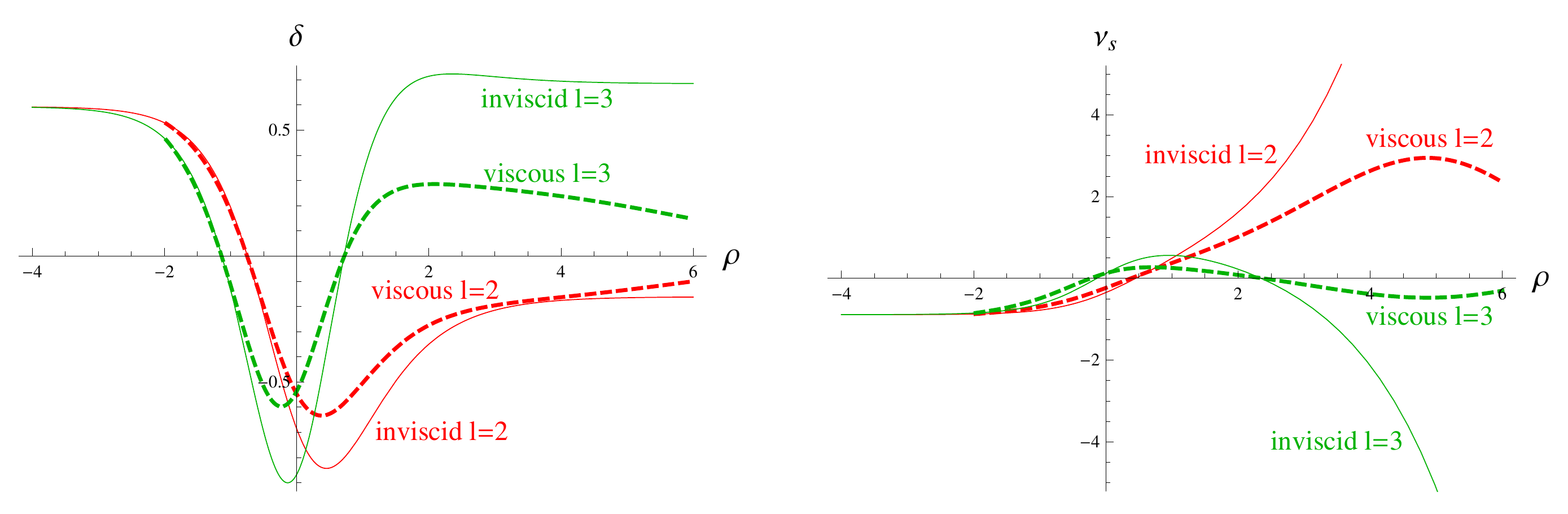}}
  \caption{(Color online) Inviscid (solid) and viscous (dashed) perturbations of the $SO(3)$-invariant flow.  The temperature perturbations $\delta$ in the inviscid case are given by \eno{deltaAgain}, and the velocity perturbations $\nu_s$ follow from \eno{uExpress}.  The viscous perturbations are obtained numerically, with initial conditions chosen at $\rho=-1.97$ (corresponding at $x_\perp=0$ to $\tau = 0.6\,{\rm fm}/c$) to agree with the inviscid perturbations.}\label{CompareViscous}
 \end{figure}

Consider first the scalar modes \eno{deltaSoln} in the inviscid theory, with $k_\eta$ set to $0$ (i.e.~the perturbation doesn't break $SO(1,1)$).  Let us also set $K_+ = 0$, so that the temperature perturbation is
 \eqn{deltaAgain}{
  \delta(\rho) = K_- (\sech\rho)^{2/3} P^{2/3}_{-{1 \over 2} + {1 \over 6}
      \sqrt{1 + 12\ell(\ell+1)}}(-\tanh\rho) \,.
 }
The point of setting $K_+=0$ is that then the velocity perturbation $\nu_s$ is as small as it can be in the early time limit.  An intriguing feature of these modes is that one can predict their amplitude at large $\rho$ through a very simple formula:
 \eqn{CosPredict}{
  R_\ell \equiv {\displaystyle\lim_{\rho \to \infty} \delta(\rho) \over 
   \displaystyle\lim_{\rho \to -\infty} \delta(\rho)} = 
    {2 \over \sqrt{3}} \cos\left( {\pi \over 6} \sqrt{1 + 12 \ell(\ell+1)} \right) 
    = \left\{ \seqalign{\span\TL &\qquad\span\TT}{
     -1 & for $\ell = 1$  \cr\noalign{\vskip-3\jot}
     -0.27 & for $\ell = 2$  \cr\noalign{\vskip-3\jot}
     1.15 & for $\ell = 3$  \cr\noalign{\vskip-3\jot}
     -0.31 & for $\ell = 4$  \cr\noalign{\vskip-3\jot}
     -1 & for $\ell = 5$ \,.
    } \right.
 }
The low odd multipoles are thus significantly enhanced relative to the low even multipoles, as measured by their propagation from $\rho = -\infty$ to $\rho = \infty$.  It might seem from this result that we are predicting an enhancement of odd multipoles over even in heavy ion collisions.  Unfortunately, this is not true, for two reasons.  First, as is clear from figure~\ref{TimeSequence}, freezeout surfaces are far from being located at large $\rho$: indeed, with the assumptions that went into this figure, the kinetic freezeout surface only barely extends to $\rho>0$.  And second, viscous corrections are significant for our usual, quasi-realistic choice of parameters \eno{SampleParams}.  The effect of viscous corrections can be seen by numerically integrating the viscous equations \eno{ScalarConserve}, still with $k_\eta=0$, and with initial conditions chosen at some reasonable initial time to be as close as possible to the inviscid form \eno{CosPredict}.  In figure~\ref{CompareViscous} we show the results of such integrations.  The effort involved in finding the viscous curves in figure~\ref{CompareViscous} is slight compared with full viscous hydrodynamics codes.

In order to gain further intuition about the perturbed conformal flows, let us now study them directly in the future wedge of ${\bf R}^{3,1}$.  We will not address hadronization, and for simplicity we will work in the inviscid theory; thus our results are not wholly realistic, and are meant mostly to convey in familiar coordinates what the perturbed conformal flows mean.  We will focus on the low multipoles of the scalar modes, namely
 \eqn{FourPerturbations}{
  \delta^{2,2}&= (\sech\rho)^{2/3} 
    P^{2/3}_{-{1 \over 2} + {1 \over 6} \sqrt{73}}(-\tanh\rho)
    \left[ -\sqrt{3 \over 8} Y_{2,2}(\theta,\phi) + {1 \over 2} Y_{2,0}(\theta,\phi) - 
     \sqrt{3 \over 8} Y_{2,-2}(\theta,\phi) \right]  \cr
   \delta^{2,1} &= (\sech\rho)^{2/3} P^{2/3}_{-{1 \over 2} + {1 \over 6} \sqrt{73}}(-\tanh\rho)
    \left[ {1 \over \sqrt{2}} Y_{2,-1}(\theta,\phi) - {1 \over \sqrt{2}}
      Y_{2,1}(\theta,\phi) \right] \cr
  \delta^{3,3} &= (\sech\rho)^{2/3} P^{2/3}_{-{1 \over 2} + {1 \over 6} \sqrt{145}}(-\tanh\rho)
    \left[ {1 \over \sqrt{2}} Y_{3,-3}(\theta,\phi) - {1 \over \sqrt{2}}
      Y_{3,3}(\theta,\phi) \right]  \cr
  \delta^{3,1} &= (\sech\rho)^{2/3} P^{2/3}_{-{1 \over 2} + {1 \over 6} \sqrt{145}}(-\tanh\rho)
    \left[ {1 \over \sqrt{2}} Y_{3,-1}(\theta,\phi) - {1 \over \sqrt{2}}
      Y_{3,1}(\theta,\phi) \right] \,,
 }
with corresponding perturbations of the velocity field \eqref{uExpress}. Our conventions for spherical harmonics are given in appendix \ref{A:Conventions}. As mentioned earlier, because $\delta^{2,2}$ involves modes with $m=\pm 2$, it should relate to elliptic flow.  We chose the particular admixture of $Y_{2,0}(\theta,\phi)$ so that $\delta^{2,2}$ as a whole would be invariant under an $SO(2)$ subgroup of $SO(3)$ which contains rotations of $S^2$ orthogonal to the one generated by $\partial_\phi$.  Both $\delta^{2,1}$ and $\delta^{2,2}$ are $\ell=2$ modes, so their $\rho$ dependence is the same. However, since $\theta$ as well as $\rho$ depends on $\tau$, the $\tau$ dependence of $\delta^{2,1}$ and $\delta^{2,3}$ are rather different. The same reasoning applies to the $\delta^{3,3}$ and $\delta^{3,1}$ modes.

\begin{figure}
  \hskip-0.2in\includegraphics[width=6.5in]{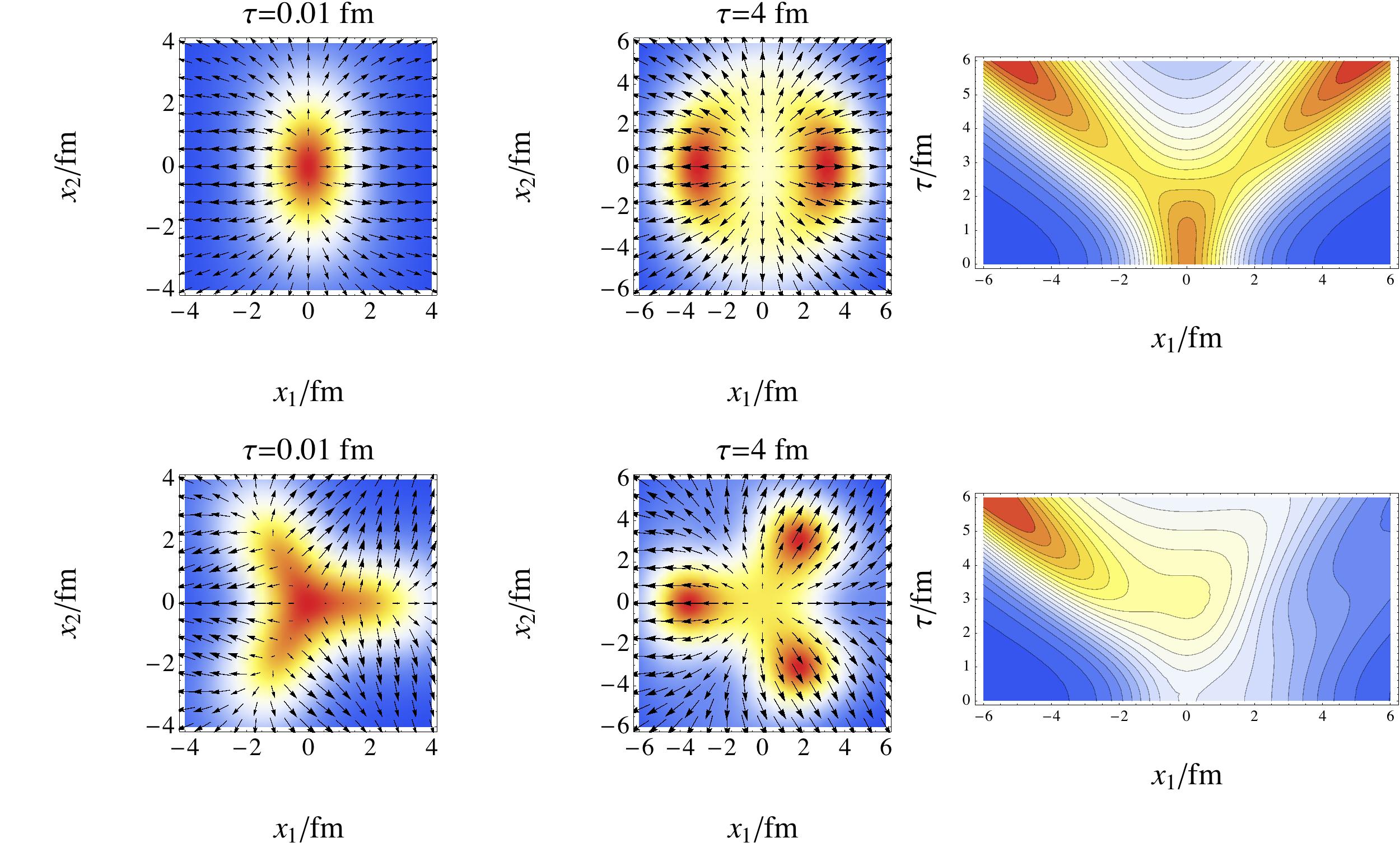}
  \caption{The energy and velocity field for the $\delta^{2,2}$ (top) and $\delta^{3,3}$ (bottom) perturbations described in equation \eqref{FourPerturbations}. The energy density is lowest in blue regions and is close to its maximum in red regions. The two left columns show density plots of the energy density overlaid with a vector plot of the velocity field whose $\eta$ component vanishes. The rightmost plots are contour plots of $\tilde{\epsilon}$ described in \eqref{tildeEpsilonDef} and provide an illustration of the time evolution of the energy density.}\label{SModesPlots1}
 \end{figure}

\begin{figure}
  \hskip-0.2in\includegraphics[width=6.5in]{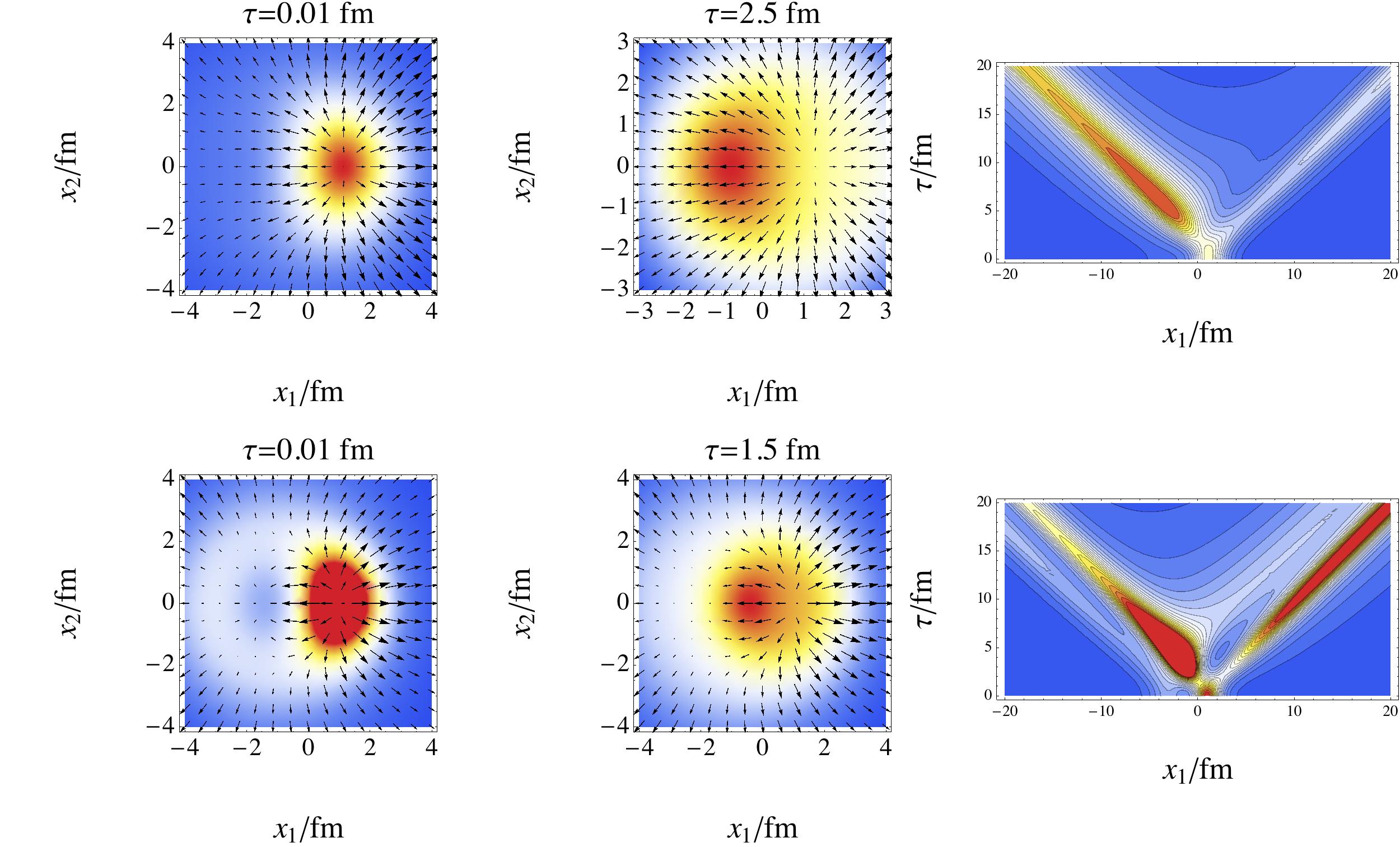}
  \caption{The energy and velocity field for the $\delta^{2,1}$ (top) and $\delta^{3,1}$ (bottom) described in equation \eqref{FourPerturbations}.  The energy density is lowest in blue regions and is close to its maximum in red regions. The two left columns show density plots of the energy density overlaid with a vector plot of the velocity field whose $\eta$ component vanishes. The rightmost plots are contour plots of $\tilde{\epsilon}$ described in \eqref{tildeEpsilonDef} and provide an illustration of the time evolution of the energy density.}\label{SModesPlots2}
 \end{figure}

Using \eno{RhoThetaCoords} one can readily express the perturbations \eno{FourPerturbations} as functions of $\tau$, $x^1$, and $x^2$.  In figures~\ref{SModesPlots1} and \ref{SModesPlots2} we show the energy density of the background flow \eno{EpsilonFlat} perturbed by the perturbations in \eqref{FourPerturbations}.
More precisely, we set $T = T_b \left( 1 + K_- \delta^{l,m} \right)$ for $\ell$ and $m$ as in \eqref{FourPerturbations}, and so the energy density plotted is $\epsilon_b \left( 1 + K_- \delta^{\ell,m} \right)^4$ with $\epsilon_b$ as in \eqref{FoundEpsilon} and $K_-=3/5$, $q=(4.3 \hbox{fm})^{-1}$ and $\hat{\epsilon}_0=880$. We have used $K_-=3/5$ in order to make the main features of the perturbation obvious. The value for $\hat{\epsilon}$ has been taken from \cite{Gubser:2010ze}.
While the temperature remains positive, the perturbations as plotted can hardly be described as small, so the linearized approximation may not be justified for such a large value of $K_-$.
For tracking the time evolution of the energy density, we found it useful to plot the quantity
 \eqn{tildeEpsilonDef}{
  \tilde\epsilon(\tau,x^1) \equiv {\epsilon(\tau,x^1,0,0) \over \epsilon_b(\tau,\tau,0,0)} = 
    {\epsilon_b(\tau,x^1,0,0) \over \epsilon_b(\tau,\tau,0,0)} (1+\delta)^4 \,.
 }
Put in words, $\tilde\epsilon$ is the perturbed energy density divided by the unperturbed energy density evaluated at $x^1 = \tau$ with $x^2 = 0$.  This quantity is convenient because its global maximum on any slice of constant Bjorken time $\tau$ is close to $1$.

\begin{figure}
  \begin{center}
   \includegraphics[width=6in]{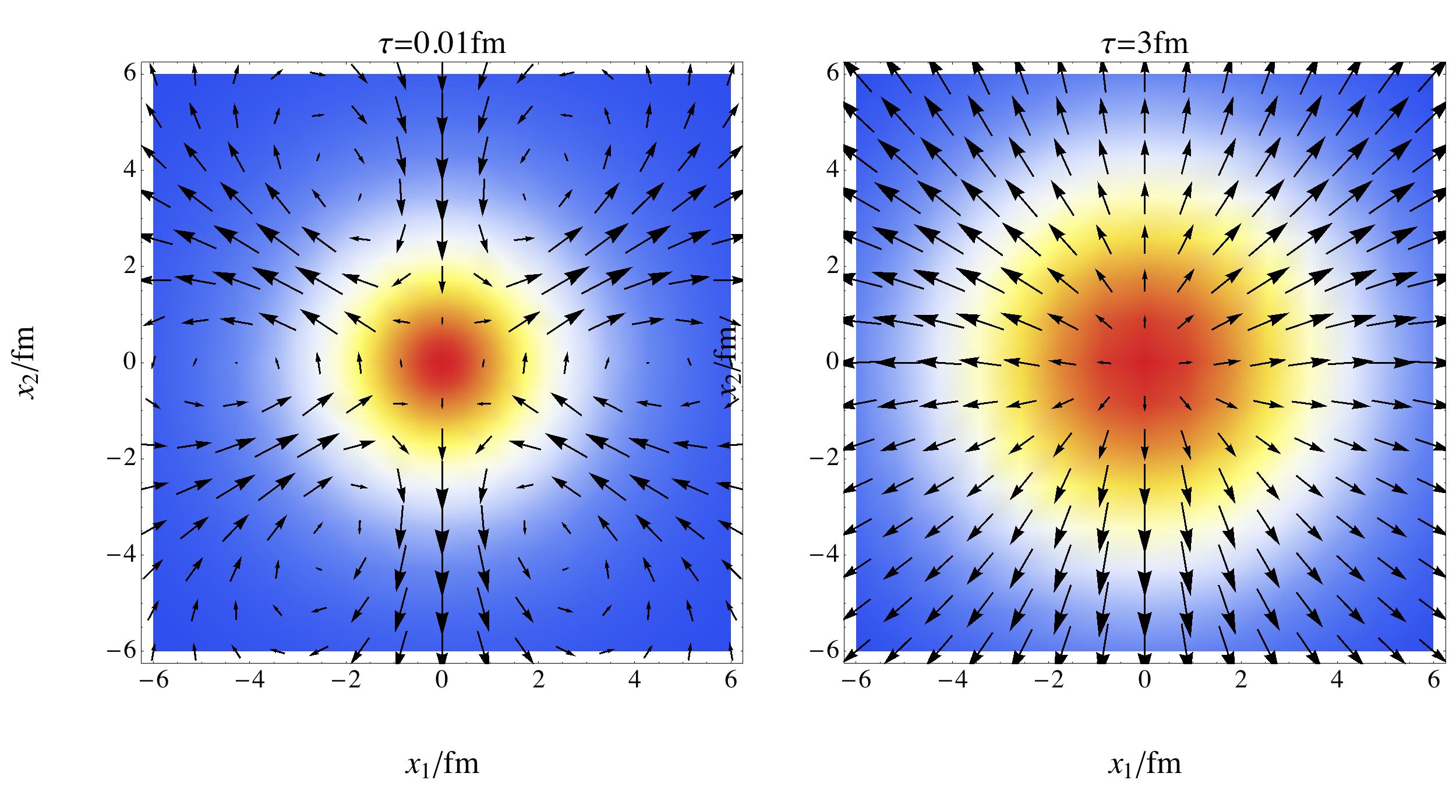} 
  \end{center}
  \caption{The energy and velocity field for the vector perturbation $V^{3,3}$ described in equation \eqref{E:V33}.  The energy density is lowest in blue regions and is close to its maximum in red regions.}\label{VModesPlot}
 \end{figure}

For illustrative purposes, we've also plotted, in figure \ref{VModesPlot}, vector perturbations of the form $\nu_i = K_V V^{3,3}$ with
\begin{equation}
\label{E:V33}
	V^{3,3} = (\cosh\rho)^{2/3} \frac{1}{\sqrt{2}}\left(\Phi_{3,-3}-\Phi_{3,3}\right)\,
\end{equation}
and $K_v = 3/5$.
As expected, the vector perturbations don't affect the rest frame energy density $\epsilon$ and exhibit a non trivial velocity field. In the lab frame, the energy density is time dependent as is the energy flux.

\subsection{Parameterizing anisotropies}
\label{PARAMETRIZE}

Recently, the authors of \cite{Alver:2010gr} have provided an explanation of certain features of two-particle correlations at RHIC which relies on the propagation of fluctuating initial conditions. Roughly, stochastic fluctuations in the initial phase of the collision evolve into anisotropic perturbations of the hydrodynamic flow which, after hadronization generate the peculiar ``ridge'' and ``shoulder'' observed in two particle correlations. In \cite{Teaney:2010vd} this idea was formulated using a cumulant expansion of the early time perturbations of the entropy density which was then fed as an initial condition to linearized hydrodynamics and finally, using a Cooper-Frye hadronization scheme, to a final hadron distribution.

It is interesting to compare our expansion of anisotropies in spherical harmonics with the cumulant expansion of \cite{Teaney:2010vd}.\footnote{
This section is based on discussions with D.~Teaney.}  To simplify the discussion, let's suppose that the stochastic anisotropies of the flow are due to anisotropies in the transverse energy density of the colliding ions which we approximate by a light-like shock wave:
 \eqn{Tuu}{
  T^v_{\phantom{v}u} = -2 f(\vec{x}_\perp) \delta(u)
 }
where $\vec{x}_\perp = (x^1,x^2)$ and 
 \eqn{uDef}{
  u = t - x^3 \qquad v = t+x^3 \,.
 }
 The actual form of the transverse distribution $f(\vec{x}_{\perp})$ is model dependent. As explained in \cite{Gubser:2010ze} (and anticipated in \cite{Gubser:2008pc,Gubser:2009sx} on AdS/CFT grounds), the unique $SO(3)$-invariant transverse distribution is
 \eqn{rhoSymmetric}{
  f_{SO(3)}(\vec{x}_\perp) = {2q^2 E \over \pi (1+q^2 x_\perp^2)^3} \,,
 }
where $E$ is the total energy of one shock wave.  A more phenomenologically motivated choice is based on the Woods-Saxon profile:
 \eqn{rhoWS}{
  f_{\rm WS}(\vec{x}_\perp) = {E \over -8\pi a^3 {\rm Li}_3(-e^{R/a})}
   \int_{-\infty}^\infty {dx^3 \over 
    1 + e^{\left[ \sqrt{x_\perp^2 + (x^3)^2} - R \right]/a}} \,,
 }
where $E$ is again the total energy and ${\rm Li}_n$ denotes the polylogarithm function.  The rms radius of the distributions \eno{rhoSymmetric} and \eno{rhoWS} are the same provided
 \eqn{qFromRa}{
  q^2 = {{\rm Li}_3(-e^{R/a}) \over 8a^2 {\rm Li}_5(-e^{R/a})} \,.
 }
Finally, following \cite{Teaney:2010vd}, we can consider
 \eqn{rhoGaussian}{
  f_{\rm Gaussian}(\vec{x}_\perp) = {q^2 E \over \pi} e^{-q^2 x_\perp^2} \,,
 }
whose rms width is the same as \eno{rhoSymmetric}.  It would be more precisely in the spirit of \cite{Teaney:2010vd} for $f$ to be the transverse density of wounded nucleons in a Glauber treatment of the early stages of a heavy ion collision.  In central collisions, this density approximately follows the transverse energy density of one of the nucleons, so the distinction between our definition of $f$ and the one used in \cite{Teaney:2010vd} is not large. In what follows we will set $q=E=1$ for simplicity.

In order to facilitate the comparison between the anisotropies parameterized by spherical harmonics in this work, and the cumulant expansion of \cite{Teaney:2010vd}, it is useful to introduce complex coordinates and derivatives in the transverse plane $(x^1,x^2)$
 \eqn{zdDefs}{\seqalign{\span\TL & \span\TR &\qquad\qquad\span\TL & \span\TR}{
  z &= x^1 + ix^2 & \bar{z} &= x^1 - i x^2  \cr
  \partial &= {1 \over 2} \left( {\partial \over \partial x^1} + {1 \over i}
    {\partial \over \partial x^2} \right) &
  \bar\partial &= {1 \over 2} \left( {\partial \over \partial x^1} - {1 \over i}
    {\partial \over \partial x^2} \right)}\,.
 }
If one takes de Sitter time $\rho \to -\infty$, corresponding to Bjorken time $\tau \to 0$ from above, then starting from the second equation in \eno{RhoThetaCoords} one can show that the transverse plane is mapped to $S^2$ as follows:
\eqn{stereographic}{
  z = e^{i\phi} \tan {\theta \over 2} \qquad 
  \bar{z} = e^{-i\phi} \tan {\theta \over 2} \,.
 }
This is the standard stereographic map, which itself is conformal.  
To find the Weyl factor $\Omega$ associated with the stereographic map \eqref{stereographic},
we note that the metric on $S^2$ takes the form
 \eqn{SymmetricMetric}{
  d\hat{s}^2 = d\theta^2 + \sin^2 \theta \, d\phi^2 = {4dzd\bar{z} \over (1+z\bar{z})^2} \,,
 }
from which we conclude that
 \eqn{OmegaStereographic}{
  \Omega(z,\bar{z}) = {1+z\bar{z} \over 2} \,.
 }

Consider the $SO(3)$-invariant transverse distribution \eno{rhoSymmetric} in complex coordinates,
\eqn{rhoComplex}{
  f_{SO(3)}(z,\bar{z}) = {\Omega(z,\bar{z})^{-3} \over 4\pi} \,.
 }
We see that $f$ maps to a constant on $S^2$ under the stereographic projection \eno{stereographic} if it is assumed to have conformal weight $3$. So for functions $f(z,\bar{z})$ not too different from $f_{SO(3)}(z,\bar{z})$, it is sensible to expand
 \eqn{rhoArbitrary}{
  f(z,\bar{z}) = \Omega(z,\bar{z})^{-3} \sum_{\ell=0}^\infty \sum_{m=-\ell}^\ell \hat{f}_{\ell,m} 
    X_{\ell,m}(z,\bar{z}) \,,
 }
where 
 \eqn{Xdef}{
  X_{\ell,m}(z,\bar{z}) = Y_{\ell,m}(\theta,\phi)
 }
with the coordinate mappings as defined in \eno{stereographic}.  It is straightforward to show that for $m \geq 0$,
 \eqn{Xdef2}{
  X_{\ell,m}(z,\bar{z}) = {z^m \over (1+z\bar{z})^\ell} P_{\ell,m}(z\bar{z}) \,,
 }
where $P_{\ell,m}(\xi)$ is a polynomial of degree $\ell-m$ with all its zeros at positive values of $\xi$.  Because $X_{\ell,m}(z,\bar{z}) = \pm X_{\ell,-m}(z,\bar{z})^*$, there is essentially no new information in the $X_{\ell,m}$ with $m<0$.  We tabulate the first few $X_{\ell,m}$ with $m \geq 0$ in Table~\ref{XandQ}.

The coefficients $\hat{f}_{\ell,m}$ can be interpreted as moments of slightly unusual quantities.  Defining
 \eqn{AnglesDef}{
  \langle A(\vec{x}_\perp) \rangle = {\int d^2 x_\perp \, f(\vec{x}_\perp) A(\vec{x}_\perp) \over
    \int d^2 x_\perp \, f(\vec{x}_\perp)} \,,
 }
for an arbitrary function $A(\vec{x}_\perp)$ with good smoothness and fall-off properties, we see that\footnote{It is straightforward to generalize \eno{rhoArbitrary} and \eno{GeneralAngle} to cases where $f$ transforms with an arbitrary weight $\alpha$ under the stereographic map: all that changes is the powers of $\Omega$.}
 \eqn{GeneralAngle}{
  \langle \Omega(z,\bar{z}) X_{\ell,m}(z,\bar{z})^* \rangle = 
    {\hat{f}_{\ell,m} \over \int d^2 x_\perp \, f(\vec{x}_\perp)} \,.
 }
Assuming that $\int d^2 x_\perp \, f(\vec{x}_\perp) = 1$ we find, with the conventions of appendix \ref{A:Conventions},
 \eqn{fCoefs}{
  \hat{f}_{1,1} = \sqrt{3 \over 8\pi} \langle \bar{z} \rangle \qquad
  \hat{f}_{1,0} = {3 \over 16\pi} \langle 1 - z\bar{z} \rangle \,.
 }
These moments can be assumed to vanish, since non-zero $\hat{f}_{1,1}$ corresponds to a displacement of the distribution in the transverse plane, and non-zero $\hat{f}_{1,0}$ corresponds to an alteration of its overall size, which can be compensated for by changing $q$.  Low-order moments which cannot be assumed to vanish include
 \eqn{MorefCoefs}{
  \hat{f}_{2,2} = \sqrt{15 \over 8\pi} \left\langle {\bar{z}^2 \over 1 + z\bar{z}}
    \right\rangle \qquad
  \hat{f}_{2,1} = \sqrt{15 \over 2\pi} \left\langle {\bar{z} \over 1+z\bar{z}}
    \right\rangle \qquad
  \hat{f}_{3,3} = \sqrt{35 \over 4\pi} \left\langle {\bar{z}^3 \over (1+z\bar{z})^2} 
    \right\rangle \,,
 }
where we have assumed $\hat{f}_{1,1}=0$ in order to simplify the expression for $\hat{f}_{2,1}$.

Now let us give a slightly simplified treatment of the cumulant expansion of \cite{Teaney:2010vd}.  The main idea is that a general function (or at least an appropriately broad class of functions) can be represented as
 \eqn{CumulantExpand}{
  f(\vec{x}_\perp) = \int {d^2 k \over (2\pi)^2}
    e^{i \vec{k}_\perp \cdot \vec{x}_\perp} e^{W(\vec{k}_\perp)} \,,
 }
where $W(\vec{k}_\perp)$ admits a Taylor series expansion.  In practice, for the scope of work considered in \cite{Teaney:2010vd}, it is generally enough to consider
 \eqn{Wform}{
  W(\vec{k}_\perp) = -{k_\perp^2 \over 4} + \delta W(\vec{k}_\perp)
 }
where $\delta W(\vec{k}_\perp)$ is treated as small.  Thus, to linear order,
 \eqn{fExpand}{
  f(\vec{x}_\perp) = \left[ 1 + \delta W\left( {1 \over i} {\partial \over \partial \vec{x}_\perp}
    \right) \right] {e^{-x_\perp^2} \over \pi} \,.
 }
Another way of expressing \eno{fExpand} is
 \eqn{fExpandAgain}{
  f(z,\bar{z}) = {e^{-z\bar{z}} \over \pi} \sum_{n=0}^\infty \sum_{m=-n \atop m+n\ \rm even}^n
    w_{n,m} Q_{n,m}(z,\bar{z})
 }
where we define
 \eqn{Qdef}{
  Q_{n,m}(z,\bar{z}) \equiv e^{z\bar{z}} 
    \partial^{n-m \over 2}\bar\partial^{n+m \over 2} e^{-z\bar{z}} \,,
 }
Clearly, the $Q_{m,n}$ are polynomials in $z$ and $\bar{z}$.  We tabulate the first few $Q_{n,m}$ with $m \geq 0$ in Table~\ref{XandQ}.  It is clear from the table that $X_{\ell,m}$ is qualitatively similar to $Q_{2\ell-|m|,m}$.  The main difference is the powers of $1+z\bar{z}$ in the denominators of the $X_{\ell,m}$, which renders them uniformly bounded across the plane.  The unboundedness of the $Q_{n,m}$ was a problem in \cite{Teaney:2010vd}, solved by an ad hoc regularization.

 \begin{table}
  \begin{center}
  \begin{tabular}{c | c c c c}
   $X_{\ell,m}\atop \rm up\ to\ a\ constant$ & $\ell=0$ & $\ell=1$ & $\ell=2$ & $\ell=3$
     \rule[-2.2ex]{0pt}{5.7ex} \\ \hline
    $m=0$ & $1$ & ${1-z\bar{z} \over 1+z\bar{z}}$ & 
    ${1-4z\bar{z}+z^2\bar{z}^2 \over (1+z\bar{z})^2}$ &
    ${1-9z\bar{z}+9z^2\bar{z}^2-z^3\bar{z}^3 \over (1+z\bar{z})^3}$  \rule[-2.2ex]{0pt}{5.7ex} \\
   $m=1$ & & ${z \over 1+z\bar{z}}$ & ${z(1-z\bar{z}) \over (1+z\bar{z})^2}$ & 
    ${z(1-3z\bar{z}+z^2\bar{z}^2) \over (1+z\bar{z})^3}$  \rule[-2.2ex]{0pt}{4.7ex} \\
   $m=2$ & & & ${z^2 \over (1+z\bar{z})^2}$ & ${z^2(1-z\bar{z}) \over (1+z\bar{z})^3}$
     \rule[-2.2ex]{0pt}{4.7ex} \\
   $m=3$ & & & & ${z^3 \over (1+z\bar{z})^3}$  \rule[-2.2ex]{0pt}{4.7ex} \\
  \end{tabular}\\[20pt]
  \begin{tabular}{c | c c c c c c}
   $Q_{n,m}\atop \rm up\ to\ a\ constant$ & $n=0$ & $n=1$ & $n=2$ & $n=3$ & $n=4$ & $n=5$ 
     \rule[-2.2ex]{0pt}{5.7ex} \\ \hline
   $m=0$ & $1$ & & $1-z\bar{z}$ & & $2-4z\bar{z}+z^2\bar{z}^2$ &  \rule[-2.2ex]{0pt}{5.7ex} \\ 
   $m=1$ & & $z$ & & $z(2-z\bar{z})$ & & 
    $z(6-6z\bar{z}+z^2\bar{z}^2)$ \rule[-2.2ex]{0pt}{4.7ex} \\
   $m=2$ & & & $z^2$ & & $z^2(3-z\bar{z})$ &  \rule[-2.2ex]{0pt}{4.7ex} \\
   $m=3$ & & & & $z^3$ & & $z^3(4-z\bar{z})$  \rule[-2.2ex]{0pt}{4.7ex} \\
  \end{tabular}
  \end{center}
  \caption{The first few $X_{\ell,m}$ and $Q_{n,m}$ with overall constant prefactors omitted.}\label{XandQ}
 \end{table}

The constants $w_{n,m}$ in \eno{fExpand} are closely related to the moments $W_{n,m}$, $W_{n,m}^c$, and $W_{n,m}^s$ of \cite{Teaney:2010vd}.\footnote{One difference in notation between the present summary and \cite{Teaney:2010vd} is that $n$ and $m$ have been switched: For us, $m$ denotes the ``magnetic quantum number'' which indicates that a function contains a factor $e^{im\phi}$, whereas in \cite{Teaney:2010vd} $n$ is the magnetic quantum number.}  To work out expressions for $w_{n,m}$ in terms of moments, it helps to define
 \eqn{kzDef}{
  k_z = {1 \over 2} (k_1-ik_2) \qquad\qquad k_{\bar{z}} = {1 \over 2} (k_1+ik_2) \,.
 }
Then from \eno{CumulantExpand}-\eno{Qdef} we read off
 \eqn{DoubleExpand}{
  1 + \delta W(k_z,k_{\bar{z}}) = \sum_{n=0}^\infty \sum_{m=-n \atop m+n\ \rm even}^n w_{n,m} 
    (ik_z)^{n-m \over 2} (ik_{\bar{z}})^{n+m \over 2}
   = e^{k_z k_{\bar{z}}} \left\langle e^{-i k_z z - i k_{\bar{z}} \bar{z}} \right\rangle \,,
 }
where we persist in treating $\delta W$ at linear order and we assume $w_{0,0} = 1$.  Matching terms in Taylor series expansions in $k_z$ and $k_{\bar{z}}$, one finds
 \eqn{wCoefs}{
  w_{1,1} = -\langle \bar{z} \rangle \qquad
   w_{2,0} = -\langle 1-z\bar{z} \rangle \,.
 }
These moments are identical to those in \eno{fCoefs}, and we again assume that they vanish.  Low-order non-vanishing moments include
 \eqn{MorewCoefs}{
  w_{2,2} = {1 \over 2} \langle \bar{z}^2 \rangle \qquad
  w_{3,1} = -{1 \over 2} \langle z\bar{z}^2 \rangle \qquad
  w_{3,3} = -{1 \over 6} \langle \bar{z}^3 \rangle \,,
 }
where we have assumed $w_{1,1}=0$ in order to simplify the expression for $w_{3,1}$.

There are two reasons in heavy-ion phenomenology to regard the moments \eno{MorefCoefs} as useful alternatives to the more commonly used ones in \eno{MorewCoefs}.  First, they are based on $SO(3)$ group content, and the strength of the perturbations that they measure would be preserved by conformally invariant dynamics.  The early time dynamics of a heavy ion collision is approximately conformal, and it seems natural to take advantage of this.  Second, purely polynomial moments like $\langle \bar{z}^3 \rangle$ have a strong bias toward surface effects.  The moment $\langle \bar{z}^3 / (1+z\bar{z})^2 \rangle$ has less surface bias.

\section{Summary}

In this work we found an analytic solution for a viscous, conformal fluid dynamics which respects an $SO(d-1)\times SO(1,1)\times \mathbf{Z}_2$ symmetry (with $d\geq 4$) which generalizes the $SO(3)$-invariant flow of \cite{Gubser:2010ze}. The $SO(1,1) \times \mathbf{Z}_2$ symmetry is the standard boost invariant symmetry often considered in the heavy ion literature, and the $SO(d-1)$ symmetry is a subgroup of the $SO(d,2)$ conformal symmetry. Our method of constructing the flow involves a Weyl rescaling and coordinate transformation of the metric which promoted conformal symmetries on ${\bf R}^{d-1,1}$ to isometries on $dS_{d-1} \times \mathbf{R}$.  In the inviscid case we also found an $SO(2)$-invariant flow in three dimensions and a general solution with a broken $\mathbf{Z}_2$ symmetry. It should be straightforward to use the same techniques to construct flows with different $SO(1,1)$ subgroups of $SO(d,2)$ i.e., flows which are not boost invariant.

When working in the $dS_{d-1} \times \mathbf{R}$ conformal frame the flow is static in the sense that the velocity field has a component only in the de-Sitter time direction. 
The simplicity of this solution allowed us to study linear perturbations around it by decomposing the perturbations into scalars and vectors on the $S^2$ in $dS_3$. In the inviscid case we found analytic expressions for the perturbation. The viscous case was studied numerically and in the short wavelength or  long time limit. An analysis of the linear perturbations revealed that the $SO(3)$-invariant flow suffers from several instabilities associated with a breaking of the $\mathbf{Z}_2$ symmetry, i.e., instabilities which generate a non-trivial velocity in the rapidity direction. Most of these instabilities are cured when the viscosity is small but non-vanishing, and the one that does remain occurs at early enough times that it is irrelevant for a study of heavy ion collisions. In short, the upshot of our analysis is the the $SO(3)$-invariant flow is stable to perturbations for parameters similar to the ones reached experimentally at RHIC.

Finally, following recent interest in higher order moments of the flow of real quark-gluon plasmas \cite{Alver:2010gr,Alver:2010dn,Petersen:2010cw,Qin:2010pf,Lacey:2010hw,Teaney:2010vd} we gave a graphical representation of our analytic expressions for the evolution of the second and third moments of the linearized scalar perturbations of the flow. In addition, we discussed how our decomposition of the modes in terms of spherical harmonics compares with the cumulant expansion of \cite{Teaney:2010vd}.

\section*{Acknowledgments}

We thank J.~Ang, E.~Shuryak, D.~Teaney, and H.~Verlinde for useful discussions. This work was supported in part by the Department of Energy under Grant No. DE-FG02-91ER40671.

\clearpage

\begin{appendix}

\section{de-Sitter space coordinates}
\label{A:deSitter}
In section \ref{SWITCHING} we discussed various coordinate systems for de Sitter space. In what follows we will provide a slightly more detailed exposition of such coordinate systems emphasizing their geometric aspects. Most of this material exists in the literature and can be found in, for example, \cite{HandL}. We have reproduced it for the sake of completeness. In what follows we discuss the $dS_3$ geometry most relevant to this work.

\begin{figure}
		\begin{center}
		\includegraphics[width=2.3 in]{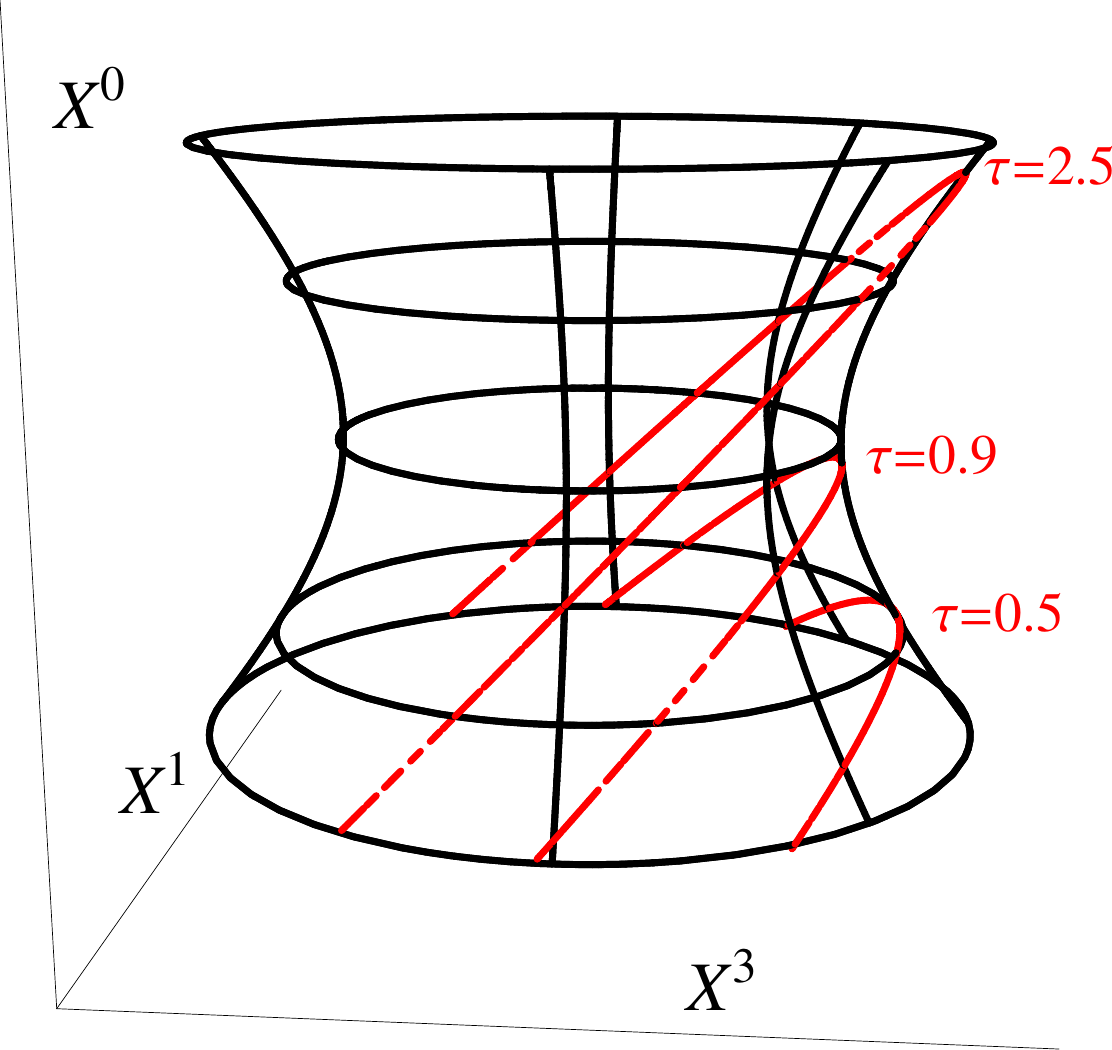}
		\caption{The $dS_3$ Hyperboloid (with the $X^2$ direction suppressed). Hypersurfaces of constant $\rho$ are two spheres, which are depicted as black circles in the figure. The radius of the $S^2$ is minimized at $\rho=0$. Lines of constant $\phi$ and $\theta$ form the boundaries of the hyperbola. The dashed red lines are surfaces of constant $\tau$}
		\label{F:hyperbola}
		\end{center}
\end{figure}

Three dimensional de Sitter space can be visualized as the hyperboloid 
\begin{equation}
	-(X^0)^2 + (X^1)^2+(X^2)^2+(X^3)^2 = L^2
\end{equation}
in $R^{3,1}$ with line element
\begin{equation}
	ds^2 = -(dX^0)^2+(dX^1)^2+(dX^2)^2+(dX^3)^2\,.
\end{equation}
In what follows we will set $L=1$ for convenience. 
The coordinate system \eqref{GlobalToCovering},
\begin{equation}
	X^0 = \sinh\rho \qquad 
	X^1 = \cosh\rho \sin\theta\cos\phi \qquad
	X^2 = \cosh\rho \sin\theta\sin\phi \qquad
	X^3 = \cosh\rho \cos\theta
\end{equation}
covers the hyperboloid. In these coordinates the line element takes the form
\begin{equation}
	ds^2 = -d\rho^2 + \cosh^2\rho \left(d\theta^2 + \sin^2 \theta d\phi^2\right)\,.
\end{equation}
The constant $\rho$ spatial hypersurfaces are two-spheres $S^2$, parameterized by $\theta$ and $\phi$, whose radius shrinks as one goes from $\rho = -\infty$ to $\rho=0$ and then expands as $\rho$ grows from $0$ to $\infty$. At $\rho=0$ the radius of the $S^2$ reaches its minimal value, $1$. See figure \ref{F:hyperbola}.

The coordinate transformation
\begin{align}
\begin{split}
	X^0 &= -\frac{1- \tau^2+ (x^1)^2 +  (x^2)^2}{2 \tau} \qquad
	X^1 = \frac{x^1}{\tau} \\
	X^3 &= \frac{1+ \tau^2 - (x^1)^2 - (x^2)^2}{2 \tau} \qquad
	X^2 = \frac{x^2}{\tau}
\end{split}
\end{align}
is identical to  \eqref{PoincareToCovering} except that we have set $q=1$ for convenience. Since
\begin{equation}
	\frac{1}{\tau} = X^3 -  X^0\,,
\end{equation}
spacelike surfaces of constant $\tau$ are surfaces of constant $X^3-X^0$ on the hyperboloid. 
Since $\tau>0$, the $(\tau,x^1,x^2)$ coordinate system covers only half of the hyperboloid. This is depicted in figure \ref{F:hyperbola}. 

Another way to understand the relation between the $(\tau,x^1,x^2)$ coordinates and the $(\rho,\theta,\phi)$ coordinates is to note that the spacelike surface of constant $\rho=\rho_0$ corresponds to 
the surface
\begin{equation}
\label{E:rrestriction}
	r^2 \equiv (x^1)^2 + (x^2)^2 = (\tau-(\cosh\rho_0-\sinh\rho_0))(\tau-(\cosh\rho_0+\sinh\rho_0))
\end{equation}
implying that $\cosh\rho_0+\sinh\rho_0 < \tau < \infty$. Implementing \eqref{E:rrestriction} in the relations
\begin{equation}
	\tan\phi = \frac{x^2}{x^1} \qquad \cot\phi = \frac{1-r^2+\tau^2}{2r}
\end{equation}
one finds that 
\begin{equation}
	0 < \phi < 2\pi \qquad
	\sinh\rho_0 < \cot\theta < \infty\,.
\end{equation}
Put differently, at time $\rho=\rho_0$, the $(\tau,x^1,x^2)$ coordinate system covers the part of the unit two-sphere given by
\begin{equation}
	(x^1)^2+(x^2)^2+(x^3)^2 = 1
	\qquad
	x^3 > \tanh\rho_0
\end{equation}

\section{Conventions}
\label{A:Conventions}
Our conventions for spherical harmonics are the standard ones implemented in Mathematica 7.0. The first few are:
\eqn{Ylms}{
   Y_{0,0}(\theta,\phi) &= \frac{1}{2 \sqrt{\pi }}  \cr
   Y_{1,-1}(\theta,\phi) &= \frac{1}{2} \sqrt{\frac{3}{2 \pi }} e^{-i \phi } \sin \theta  \cr
   Y_{1,0}(\theta,\phi) &= \frac{1}{2} \sqrt{\frac{3}{\pi }} \cos \theta  \cr
   Y_{1,1}(\theta,\phi) &= -\frac{1}{2} \sqrt{\frac{3}{2 \pi }} e^{i \phi } \sin \theta  \cr
   Y_{2,-2}(\theta,\phi) &= \frac{1}{4} \sqrt{\frac{15}{2 \pi }} e^{-2 i \phi } \sin ^2\theta  \cr
   Y_{2,-1}(\theta,\phi) &= \frac{1}{2} \sqrt{\frac{15}{2 \pi }} e^{-i \phi } \sin \theta \cos \theta  \cr
   Y_{2,0}(\theta,\phi) &= \frac{1}{4} \sqrt{\frac{5}{\pi }} \left(3 \cos ^2\theta-1\right)  \cr
   Y_{2,1}(\theta,\phi) &= -\frac{1}{2} \sqrt{\frac{15}{2 \pi }} e^{i \phi } \sin \theta \cos \theta  \cr
   Y_{2,2}(\theta,\phi) &= \frac{1}{4} \sqrt{\frac{15}{2 \pi }} e^{2 i \phi } \sin ^2\theta  \cr
   Y_{3,-3}(\theta,\phi) &= \frac{1}{8} \sqrt{\frac{35}{\pi }} e^{-3 i \phi } \sin ^3\theta  \cr
   Y_{3,-2}(\theta,\phi) &= \frac{1}{4} \sqrt{\frac{105}{2 \pi }} e^{-2 i \phi } \sin ^2\theta \cos \theta  \cr
   Y_{3,-1}(\theta,\phi) &= \frac{1}{8} \sqrt{\frac{21}{\pi }} e^{-i \phi } \sin \theta \left(5 \cos ^2\theta-1\right)  \cr
   Y_{3,0}(\theta,\phi) &= \frac{1}{4} \sqrt{\frac{7}{\pi }} \left(5 \cos ^3\theta-3 \cos \theta\right)  \cr
   Y_{3,1}(\theta,\phi) &= -\frac{1}{8} \sqrt{\frac{21}{\pi }} e^{i \phi } \sin \theta \left(5 \cos ^2\theta-1\right)  \cr
   Y_{3,2}(\theta,\phi) &= \frac{1}{4} \sqrt{\frac{105}{2 \pi }} e^{2 i \phi } \sin ^2\theta \cos \theta  \cr
   Y_{3,3}(\theta,\phi) &= -\frac{1}{8} \sqrt{\frac{35}{\pi }} e^{3 i \phi } \sin ^3\theta\,.
 }

\end{appendix}

\clearpage

\bibliographystyle{ssg}
\bibliography{map}

\end{document}